\documentclass[aps,prd,reprint,superscriptaddress,nofootinbib,showpacs,preprintnumbers]{revtex4-1}

\usepackage{graphicx} 
                
\usepackage{subfigure}
\usepackage{longtable}
\usepackage{multirow}
\usepackage{textcomp}
\usepackage{units}
\usepackage{amssymb}
\usepackage{amsmath}
\usepackage{hyperref}

\def\ra{\ensuremath{\rightarrow}}

\def\antibar#1{\ensuremath{#1\bar{#1}}}

\def\ttbar{\antibar{t}}

\def\Zboson{\ensuremath{Z}}

\def\Wboson{\ensuremath{W}}%

\def\pt{\ensuremath{p_{\mathrm{T}}}} 
\def\pT{\ensuremath{p_{\mathrm{T}}}} 
\def\et{\ensuremath{E_{\mathrm{T}}}} 
 
\def\ET{\ensuremath{E_{\mathrm{T}}}}

\def\Zmm{\ensuremath{Z \rightarrow \mu\mu}}
\def\Zee{\ensuremath{Z \rightarrow ee}}
\def\Zll{\ensuremath{Z \rightarrow \ell\ell}}
\def\Wln{\ensuremath{W \rightarrow \ell\nu}}
\def\Wen{\ensuremath{W \rightarrow e\nu}}
\def\Wmn{\ensuremath{W \rightarrow \mu\nu}}
\def\MET{\ensuremath{E_{\mathrm{T}}^{\mathrm{miss}}}}
\def\met{\ensuremath{E_{\mathrm{T}}^{\mathrm{miss}}}} 
\def\Ztau{\ensuremath{Z \rightarrow \tau\tau}}
\def\Wtau{\ensuremath{W \rightarrow \tau\nu}}
\def\TeV{\ifmmode {\mathrm{\ Te\kern -0.1em V}}\else
                   \textrm{Te\kern -0.1em V}\fi}%
\def\GeV{\ifmmode {\mathrm{\ Ge\kern -0.1em V}}\else
                   \textrm{Ge\kern -0.1em V}\fi}%
\def\MeV{\ifmmode {\mathrm{\ Me\kern -0.1em V}}\else
                   \textrm{Me\kern -0.1em V}\fi}%
\def\keV{\ifmmode {\mathrm{\ ke\kern -0.1em V}}\else
                   \textrm{ke\kern -0.1em V}\fi}%
\def\eV{\ifmmode  {\mathrm{\ e\kern -0.1em V}}\else
                   \textrm{e\kern -0.1em V}\fi}%

\let\gev=\GeV

\def\rts {\ensuremath{\sqrt{s}}}

\newcommand{\Zmumu}{\ensuremath{Z \ra \mu \mu}}

\newcommand{\mT}{\ensuremath{m_\mathrm{T}}}
\newcommand{\mt}{\ensuremath{m_\mathrm{T}}}
\newcommand{\mvis}{\ensuremath{m_\mathrm{vis}}}

\newcommand{\Ztautauemu}{\ensuremath{\Ztau \rightarrow e\mu + 4\nu}}
\newcommand{\Ztautaumumu}{\ensuremath{\Ztau \rightarrow \mu\mu + 4\nu}}

\newcommand{\sumetmet}{\ensuremath{\sum\et + \met}}

\newcommand{\muhadchan}{\ensuremath{\tau_{\mu}\tau_{h}}}
\newcommand{\ehadchan}{\ensuremath{\tau_{e}\tau_{h}}}
\newcommand{\emuchan}{\ensuremath{\tau_{e}\tau_{\mu}}}
\newcommand{\mumuchan}{\ensuremath{\tau_{\mu}\tau_{\mu}}}

\begin{document}
\preprint{CERN-PH-EP-2011-097}
\preprint{Submitted to Physical Review D}

\vspace*{0.2cm}
\title{Measurement of the {\boldmath$\ensuremath{Z}\rightarrow \tau \tau$} Cross Section with the ATLAS Detector}

\author{The ATLAS Collaboration}

\thanks{Full author list given at the end of the article.}

\noaffiliation

\date{October 24, 2011}

\begin{abstract}
The  $\ensuremath{Z} \rightarrow \tau \tau$ cross section is measured
with the ATLAS experiment at the LHC in four different 
final states determined by the decay modes of the $\tau$ leptons:
muon-hadron, electron-hadron,  electron-muon, and muon-muon.
The analysis is based on a data sample corresponding to an integrated
luminosity of $36$~pb$^{-1},$ at a proton-proton center-of-mass energy of $\sqrt{s} = 7~\TeV$.
Cross sections are measured separately for each final state in fiducial regions of high detector acceptance, 
as well as in the full phase space, over the mass region $66 - 116 \GeV$. The individual cross sections are  combined and 
the product of the total $Z$ production cross section and $\ensuremath{Z} \rightarrow \tau\tau$ 
branching fraction 
 is measured to be $0.97 \pm 0.07\mbox{ (stat)} \pm 0.06\mbox{ (syst)} \pm 0.03\mbox{ (lumi) nb}$,
 in agreement with NNLO calculations.
\end{abstract}


\maketitle
\section{Introduction}
\label{sec:introduction}

Tau leptons play a significant role in the search for new physics phenomena
at CERN's Large Hadron Collider (LHC). Hence decays of Standard Model gauge bosons to $\tau$ leptons, 
$\ensuremath{W}\rightarrow\tau\nu$ and $\ensuremath{Z}\rightarrow \tau \tau$,  
are important background processes in such searches and their production cross sections need
to be measured precisely. Studies of $\ensuremath{Z} \rightarrow \tau \tau$ processes
at the LHC center-of-mass energies are also interesting in their own right, complementing the measurements 
of the \Zboson\ boson through the electron and muon decay modes. 
Finally, measuring the cross section of a well-known Standard Model process involving $\tau$ leptons is  
highly important for the commissioning and validation of $\tau$ identification techniques, which will be crucial 
for fully exploiting the ATLAS experiment's potential in searches for  new physics involving $\tau$ leptons.

This paper describes the measurement of the \Ztau\ cross section, using four different final states and
an integrated luminosity of $36$~pb$^{-1},$
in $pp$ collisions at a center-of-mass energy of $\rts = 7~\TeV$\ recorded with the ATLAS detector~\cite{DetectorPaper:2008} at the LHC.
Two of the considered final states are the semileptonic modes  
$\ensuremath{Z} \ra  \tau\tau \ra \mu  + \textrm{hadrons} + 3\nu$ ($\tau_\mu \tau_h$) and $\ensuremath{Z} \ra \tau\tau \ra e + \textrm{hadrons} + 3\nu$ ($\tau_e \tau_h $) with branching fractions
 $(22.50 \pm 0.09)$\% and $(23.13 \pm 0.09)$\%  respectively~\cite{pdgBook}.
The remaining two final states are the leptonic modes \Ztautauemu~ ($\tau_e \tau_\mu$) and  \Ztautaumumu~($\tau_\mu \tau_\mu$)
with branching fractions $(6.20 \pm 0.02)$\% and $(3.01 \pm 0.01)$\%, respectively~\cite{pdgBook}.
Due to the large expected multijet background contamination, the $\tau_h \tau_h $ and $\tau_e \tau_e $ final states are 
not considered in this publication.
   
The  $\ensuremath{Z} \rightarrow \tau \tau$ cross section  has been measured previously in $p\bar{p}$ collisions at the Tevatron using the semileptonic $\tau$ decay modes~\cite{D0,CDF}. 
More recently the cross section, using both the semileptonic and leptonic modes, was
measured in $pp$ collisions at the LHC by the CMS Collaboration~\cite{CMStautau}.

After a brief description of the ATLAS detector in Section \ref{sec:atlas}, 
the data and Monte Carlo samples are presented in Section \ref{sec:data_samples}.
The object and event selections are detailed in Section \ref{sec:event_selection}.
The estimation of the backgrounds is described in 
Section \ref{sec:background_estimation}. The calculation of the cross sections is outlined in Section \ref{sec:azcz}, and a discussion of the systematic uncertainties is given in Section \ref{sec:systematics}. 
The  results, including the combination of the four channels, are presented in Section \ref{sec:results}. 

\section{The ATLAS detector}
\label{sec:atlas}

The ATLAS detector~\cite{DetectorPaper:2008}  is a multi-purpose apparatus operating at the LHC, 
designed to study a range of physics processes as wide as possible. 
ATLAS consists of several layers of subdetectors -- from the interaction point outwards, 
the inner detector tracking system, the
electromagnetic and hadronic calorimeters, and  
the muon system.\footnote{ATLAS uses a right-handed coordinate system with its origin at the nominal interaction point (IP) in the center of the detector and the $z$-axis along the beam pipe. The $x$-axis points from the IP to the center of the LHC ring, and the $y$ axis points upward. Cylindrical coordinates $(r,\phi)$ are used in the transverse plane, $\phi$ being the azimuthal angle around the beam pipe. The pseudorapidity is defined in terms of the polar angle $\theta$ as $\eta=-\ln\tan(\theta/2)$.  The distance~$\Delta R$ in the $\eta-\phi$ space is defined as 
$\Delta R = \sqrt{(\Delta\eta)^2+(\Delta\phi)^2}$.}

The inner detector is immersed in a  2~T magnetic field generated by the central solenoid.
It is designed to provide high-precision tracking information for charged particles 
and consists of three subsystems, the Pixel detector, the Semi-Conductor Tracker (SCT), and the Transition 
Radiation Tracker (TRT). The first two subsystems cover a region of $|\eta|<2.5$ in pseudorapidity, while the TRT reaches up to $|\eta|=2.0$. A track in the barrel region typically produces 11 hits in the Pixel and SCT detectors and 36 hits in the TRT.

The electromagnetic (EM) and hadronic calorimeters  cover the range $|\eta|<4.9$, with the $\eta$ region matched to  
the inner detector having a finer granularity in the EM section, needed for precision measurements of electrons and photons. 
The EM calorimeter uses lead as an absorber and
liquid argon (LAr) as the active material. The hadronic calorimeter uses steel and scintillating tiles in the barrel 
region, while the end-caps use LAr as the active material and copper as the absorber. 
The forward calorimeter also uses LAr as the active medium with copper and tungsten absorbers.

The muon spectrometer relies on the deflection of muons as they pass through the magnetic field of the 
large superconducting air-core toroid magnets. 
The precision measurement of muon track coordinates in the bending direction of the magnetic field is provided, 
over most of the $\eta$-range, by Monitored Drift Tubes (MDT).
Cathode Strip Chambers (CSC) are used in the innermost plane for $2.0<|\eta|<2.7$ due to the high particle rate in that region. 
The muon trigger, as well as the coordinate in the direction orthogonal to the bending plane,
are provided by Resistive Plate Chambers (RPC) in the barrel and Thin Gap Chambers (TGC) in the end-caps.

The ATLAS detector has a three-level trigger system consisting of Level-1 (L1), Level-2 (L2), and the
Event Filter (EF).

At design luminosity the L1 trigger rate is approximately 75 kHz. The L2 and EF triggers reduce the event rate to 
approximately 200 Hz before data transfer to mass storage.

\section{Data and Monte Carlo Samples}
\label{sec:data_samples}

The data sample used in this analysis corresponds to a total integrated luminosity 
of about $36$ pb$^{-1}$, recorded with stable beam conditions and a fully operational detector in 2010. 

Events are selected using either single-muon or single-electron triggers with thresholds based 
 on the transverse momentum ($\pt$) or transverse energy ($\et$) of the muon or electron candidate, respectively. For the \muhadchan\ and \mumuchan\ final states, single-muon triggers requiring  $\pt > 10-13~\GeV$, depending on the run period, are used. For the \ehadchan\ and \emuchan\ final states, a single-electron trigger requiring $\et > 15~\GeV$\ is used. In the  \emuchan\ final state the choice was made to use a single-electron trigger rather than a single-muon trigger because it is more efficient, as well as allowing a low offline $\pt$ cut on the muon.

The efficiency for the muon trigger is determined from data using the so-called ``tag-and-probe method'',  applied to \Zmm\ events.
It is found to be close to 95\% in the end-cap region, and around 80\% in the barrel region (as expected from the geometrical
coverage of the RPC).

Similarly, the electron trigger efficiency is measured in data, using \Wen\ and \Zee\ events. It is measured to be $\sim 99$\% for offline electron candidates with $\et>20$~\GeV\ 
and $\sim 96$\% for electron candidates with $\et$ between 16 and 20 \GeV~\cite{Triggerpaper}.

The signal and background Monte Carlo (MC) samples used for this study are generated at $\sqrt{s}=7$~\TeV~with 
the default ATLAS MC10 tune~\cite{MC10} and passed through a full detector simulation based on 
the \textsc{geant4} program~\cite{geant4,*simulation}. 
The inclusive $W$ and $\gamma^{*}/Z$ signal and background samples are generated 
with \textsc{pythia 6.421}~\cite{pythia} and are normalized to NNLO cross sections~\cite{fewz_Melnikov, *fewz_Gavin,*Catani}.

For  the $\ttbar$ background the \textsc{mc$@$nlo} generator is used~\cite{Frixione:2002}, while the diboson samples are generated with 
\textsc{herwig}~\cite{herwig}. 

In all samples the $\tau$ decays are modeled with \textsc{tauola}~\cite{TauolaPaper}.
All generators are interfaced to \textsc{photos}~\cite{PhotosPaper} to simulate the effect of final state QED radiation.

\section{Selection of $\ensuremath{Z}\rightarrow \tau \tau$ candidates}
\label{sec:event_selection}

The event preselection selects events containing at least one primary vertex with three or more associated tracks, as 
well as aiming to reject events with jets or $\tau$ candidates caused by out-of-time cosmic-rays events or known noise effects in the
calorimeters.

In the case of the two semileptonic decay modes, events are characterized by the presence 
of an isolated lepton\footnote{In the following, the term ``lepton'', $\ell$, refers to electrons and muons only.} and a hadronic $\tau$ decay\footnote{In the following, reconstructed jets identified as hadronic $\tau$ decays are referred to as 
``$\tau$ candidates'' or $\tau_h$.}.  
The latter produces a highly collimated jet in the detector consisting of an
odd number of charged hadrons and additional
calorimetric energy deposits from possible $\pi^{0}$ decay products.
The two leptonic decay modes are characterized by two isolated leptons
of typically lower transverse momentum than those in $Z \rightarrow ee/\mu\mu$ events. 
Finally, in all four channels missing energy is expected from the neutrinos produced in the $\tau$ decays.
This analysis depends therefore on many reconstructed objects: electrons, muons, $\tau$ candidates, jets and
missing transverse momentum, \met.

\subsection{Reconstructed Physics Objects}
\label{sec:rec_obj}

\subsubsection{Muons}
Muon candidates are formed by associating muon spectrometer tracks with inner detector tracks
after accounting for energy loss in the calorimeter~\cite{WZcross}.
A combined transverse momentum is determined using a statistical combination of the two tracks
and is required to be greater than $15\GeV$ for the $\tau_{\mu}\tau_{h}$ final states
and $10\GeV$ for the $\tau_{e}\tau_{\mu}$ and $\tau_{\mu}\tau_{\mu}$ final states.
Muon candidates are also required to have $|\eta| < 2.4$ and a longitudinal impact parameter of less than 10~mm with respect to the primary vertex.
In the final muon selection, the combined muon tracks are also required to pass several 
inner detector track quality criteria~\cite{ATLAS-CONF-2011-063}, resulting in an 
efficiency of $\sim 92$\%, as measured in data using \Zmm~events.

\subsubsection{Electrons} 
Electron candidates are reconstructed from clusters in the EM calorimeter matched to tracks in the inner detector. Candidate electrons are selected if they have a transverse energy $\ET > 16\GeV$ and are within the rapidity range
$|\eta|<2.47$, excluding the transition region, $1.37<|\eta|<1.52$,  between the barrel and end-cap calorimeters.
For the  $\tau_{e}\tau_{\mu}$ final state, the candidates are required to pass the  ``medium'' identification requirements based on the calorimeter 
shower shape, track quality, and track matching with the calorimeter cluster as described in~\cite{WZcross}.
The resulting efficiency is $\sim 89\%$. For the $\tau_{e}\tau_{h}$ 
final state, the electron candidate is instead required to pass the ``tight''  identification criteria, with an efficiency of 
$\sim 73$\%. 
In addition to the ``medium'' criteria, the ``tight'' selection places more stringent requirements on the
track quality, the matching of the track to the calorimeter cluster, the ratio between the calorimeter energy
and the track momentum, and the transition radiation in the TRT~\cite{WZcross}.
The electron reconstruction and identification efficiencies are measured in data using $W\ra e\nu$ and $Z \ra ee$ events. 

\subsubsection{Jets} 
The jets used in this analysis are reconstructed using the anti-$k_\mathrm{T}$ algorithm~\cite{ANTI_KT,*ANTI_KTweb},
with a distance parameter $R = 0.4$, using three-dimensional topological calorimeter energy clusters
as inputs. The energy of the jets  is calibrated using $\pt$ and  $\eta$-dependent correction factors~\cite{jet_calib}
based on simulation and validated by test beam and collision data.
Jet candidates are required to have a transverse
momentum $\pt > 20 \GeV$ and a rapidity within $|\eta|<4.5$. 
        
\subsubsection{Hadronic $\tau$ candidates} 
The reconstruction of hadronic $\tau$ decays is seeded by calorimeter jets. Their energy is determined by
applying a MC-based correction to the
reconstructed energy in the calorimeters.
Tracks with $\pt > 1\GeV$ passing minimum quality criteria are associated to calorimeter jets to form $\tau$ candidates.
Reconstructed $\tau$ candidates are selected 
if they have a transverse momentum  $\pt > 20 \GeV$ and lie 
within the pseudorapidity range $|\eta|<2.47$, excluding the 
calorimeter transition region, $1.37<|\eta|<1.52$.
Further, a $\tau$ candidate is  required to pass identification selection criteria,
based on three variables describing its energy-weighted transverse width in the electromagnetic calorimeter ($R_{EM}$), its $p_T$-weighted track width ($R_{track}$), and the fraction of the candidate's transverse momentum carried by the leading track.
In order to account for the increasing collimation of the $\tau$ candidates
with increasing $\pt$, the selection criteria on the quantities $R_{EM}$ and $R_{track}$ are parametrized as a function of the $\pt$ of the $\tau$
candidate. The identification is optimized separately for candidates with
one or multiple tracks. Additionally, a dedicated selection to reject fake $\tau$ candidates 
from electrons is applied.
This leads to an efficiency of $\sim 40$\% ($\sim 30$\%) for real 1 prong (3 prong) $\tau$ candidates as determined from signal Monte Carlo~\cite{ATLAS-CONF-2011-057}. For fakes from
multijet final states the efficiency is $\sim 6$\% ($\sim 2$\%) for 1 prong (3 prong) candidates, as measured in data using a dijet selection~\cite{ATLAS-CONF-2011-113}.
 
\subsubsection{Missing transverse momentum}
The missing transverse momentum (\MET) reconstruction used in all final states relies on energy deposits in the calorimeter 
and on reconstructed muon tracks. It is defined as the vectorial sum  $\met = \met(\mathrm{calo}) + \met(\mathrm{muon}) - \met(\mathrm{energy\ loss}) $, where
$\met(\mathrm{calo})$ is calculated from the energy deposits in calorimeter cells inside three-dimensional 
topological clusters~\cite{jet_calib}, 
$\met(\mathrm{muon})$ is the vector sum of the muon momenta, and 
$\met(\mathrm{energy\ loss})$ is a correction term accounting for the energy lost by muons in the calorimeters.
There is no direct requirement on \MET\ applied in this analysis but the quantity and its direction 
is used in several selection criteria described later.

\begin{figure}
  \centering
    \subfigure[]{
      \label{fig:muonisolationET}
      \includegraphics[width=0.48\textwidth]{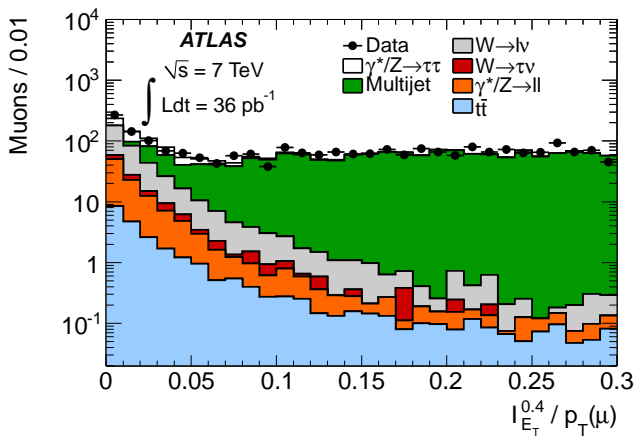}
    }
    \subfigure[]{
      \includegraphics[width=0.48\textwidth]{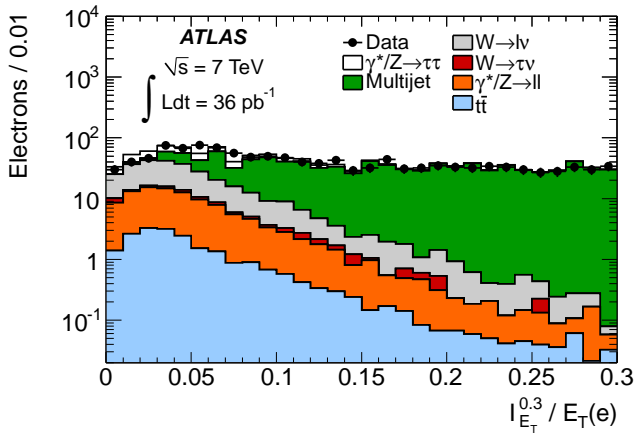}         
      \label{fig:eleisolationET}
    }
    \caption{Isolation variables (a) $\mathit{I}_{ET}^{0.4} / \pt$ for muon and (b) $\mathit{I}_{ET}^{0.3}/ \et$    for electron candidates, 
    after selecting one hadronic $\tau$ candidate and one lepton with opposite signs in $\tau_{\mu} \tau_{h}$ 
    and $\tau_{e}\tau_{h}$ final states respectively. The multijet background is estimated 
    from data according to the method described in Section \ref{sec:background_estimation}; 
    all other processes are estimated using MC simulations.
    \label{fig:lephadiso_distributions}}
\end{figure}

\subsubsection{Lepton isolation}
\label{sec:lepton_isolation}

Leptons from $\gamma^*/Z\rightarrow \tau \tau$ decays are typically isolated from other particles, in contrast to electrons 
and muons from multijet events (e.g. coming from b-hadron decays).
Hence isolation requirements are applied to both the electron and muon candidates used in the four final states considered.

The first isolation variable is based on the total transverse momentum of charged 
particles in the inner detector in a cone of size $\Delta R=0.4$ centered around the lepton direction, 
$\mathit{I}_{PT}^{0.4}$,
divided by the transverse momentum or energy of the muon or electron candidate, 
respectively.
A selection requiring $\mathit{I}_{PT}^{0.4} / \pt < 0.06$ for the muon 
candidate and $\mathit{I}_{PT}^{0.4} / \et < 0.06$ for the electron candidate  
is used for all final states except the $\tau_{\mu}\tau_{\mu}$ final state 
where a looser selection, $\mathit{I}_{PT}^{0.4} / \pt < 0.15$,
is applied. 
Due to the presence of two muon candidates the multijet background is smaller in this final state, and the looser isolation requirement provides a larger signal efficiency.

A second isolation variable is based on the total transverse energy measured in the calorimeters in a cone $\Delta R$ around the lepton
direction, $\mathit{I}_{ET}^{\Delta R}$, divided 
by the transverse momentum  or energy of the muon or electron candidate, 
respectively.
For muon candidates, a cone of size $\Delta R=0.4$ is used, and the requirement
$\mathit{I}_{ET}^{0.4} / \pt<0.06$ is applied to all final states but the $\tau_{\mu}\tau_{\mu}$ final state
where a looser selection, $\mathit{I}_{ET}^{0.4} / \pt < 0.2$, is applied.
For electron candidates, a cone of size $\Delta R=0.3$ is used and  a selection requiring $\mathit{I}_{ET}^{0.3} / \et < 0.1$ 
is applied  in both $\tau_{e}\tau_{h}$ and  $\tau_{e}\tau_{\mu}$ final states.
In the reconstruction of all the isolation variables, the lepton $\pt$ or $\et$ is subtracted.

The efficiencies for these isolation requirements are measured in data using \Zmumu\ and \Zee\ events and found to be $75-98\%$ for muons and $60-95\%$ for electrons, depending on the transverse momentum or energy respectively. 
Figure~\ref{fig:lephadiso_distributions} shows the distribution of the 
$\mathit{I}_{ET}^{0.4} / \pt$ variable for 
muon and $\mathit{I}_{ET}^{0.3} / \et$ variable for electron candidates.

\subsection{Event selection}

To select the required event topologies, the following selections are applied for the final states considered 
in this analysis:

\begin{itemize}
\item $\tau_{\mu}\tau_{h}$: at least one isolated muon candidate with $\pt > 15\GeV$ and one hadronic $\tau$ candidate
                           with $\pt > 20\GeV$,
\item $\tau_{e}\tau_{h}$: at least one isolated ``tight'' electron candidate with $\et > 16\GeV$ and one hadronic $\tau$ candidate with $\pt > 20\GeV$,
\item $\tau_{e}\tau_{\mu}$: exactly one isolated ``medium'' electron candidate with $\et > 16\GeV$ and one isolated muon candidate with $\pt > 10\GeV$,
\item $\tau_{\mu}\tau_{\mu}$: exactly two isolated muon candidates with $\pt > 10\GeV$, at least one of which should have $\pt > 15\GeV$. 
\end{itemize}
These selections are followed by a number of event-level selection criteria optimized to suppress electroweak backgrounds.

\begin{figure*}
    \centering
    \subfigure[$\tau_{\mu}\tau_{h}$ final state]{
        \includegraphics[width=0.47\textwidth]{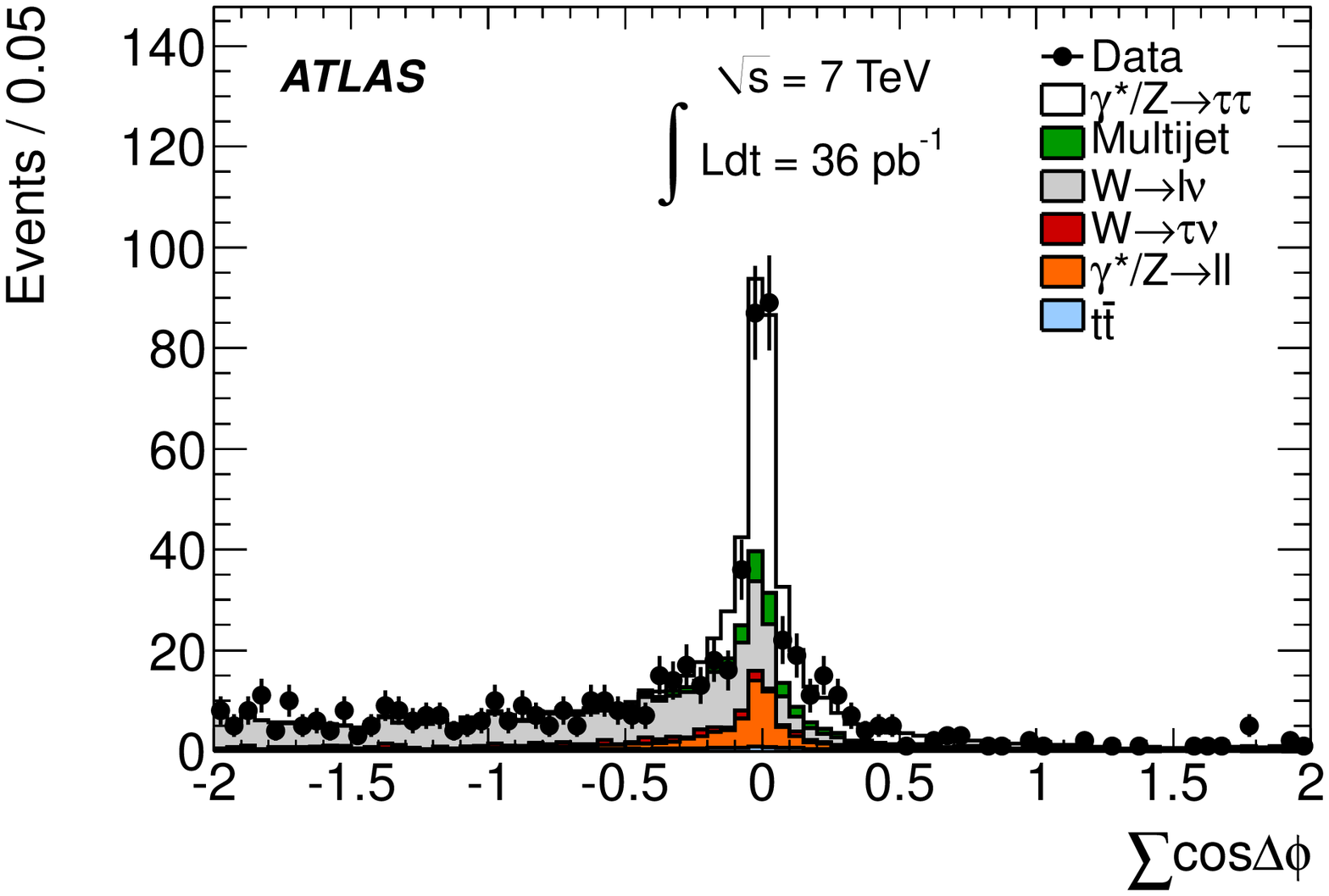}
        \label{fig:lephad_muon_h_sum_cos_dphi}
    }   
    \subfigure[$\tau_{e}\tau_{h}$ final state]{
       \includegraphics[width=0.47\textwidth]{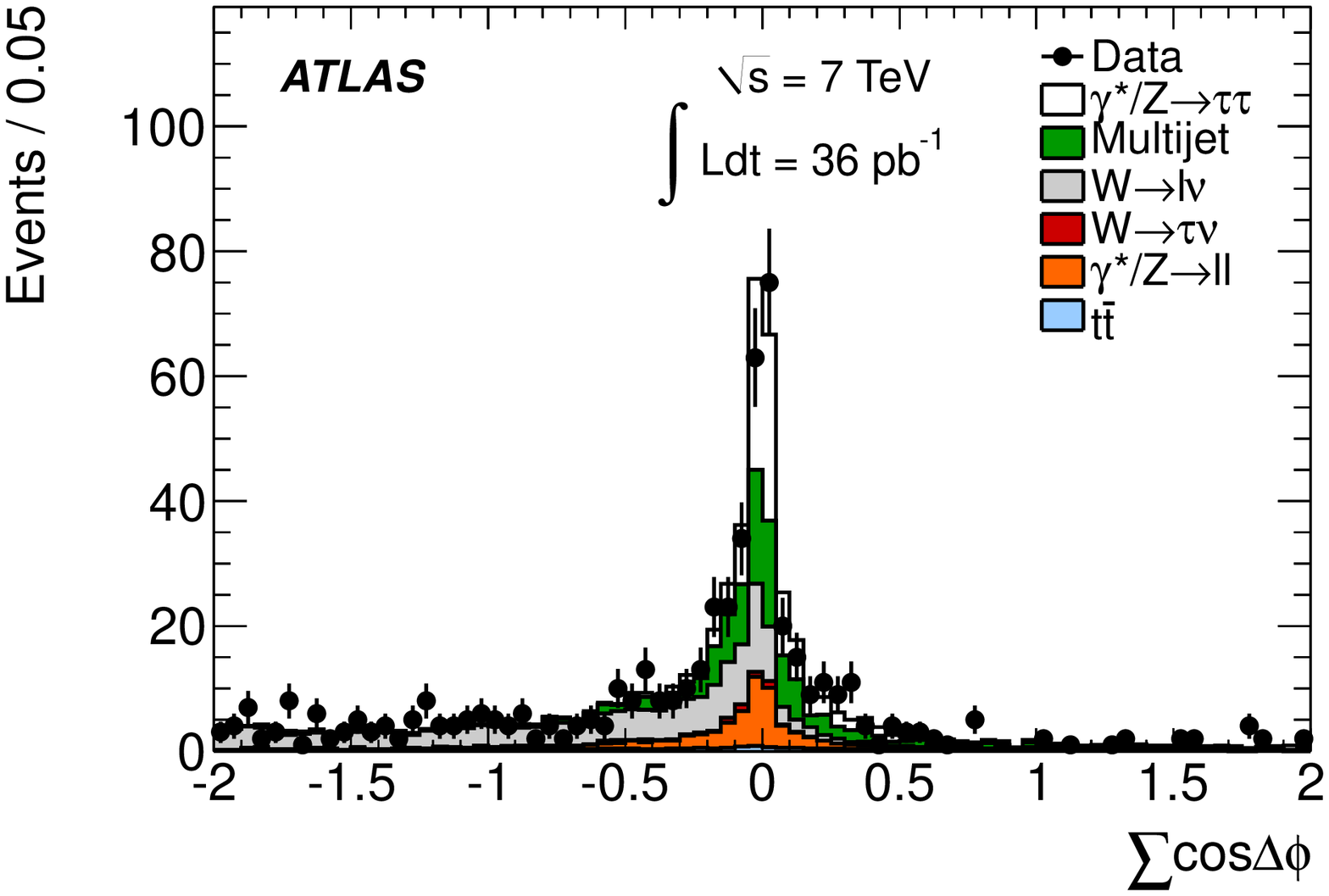}
        \label{fig:lephad_ele_h_sum_cos_dphi}
    }\\*   
    \subfigure[$\tau_{\mu}\tau_{h}$ final state]{
       \includegraphics[width=0.47\textwidth]{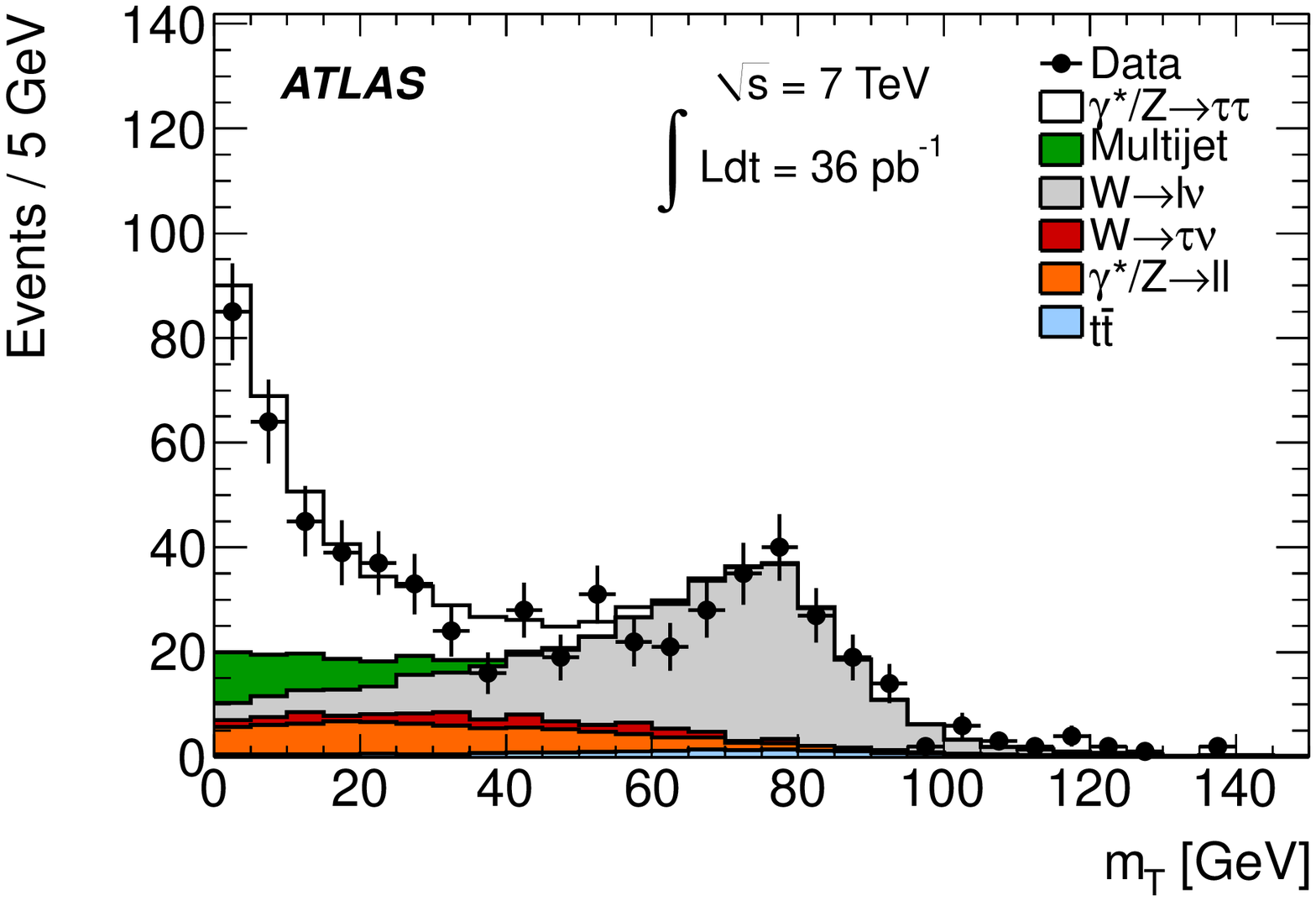}
        \label{fig:lephad_muon_h_trans_mass}
    }
    \subfigure[$\tau_{e}\tau_{h}$ final state]{
       \includegraphics[width=0.47\textwidth]{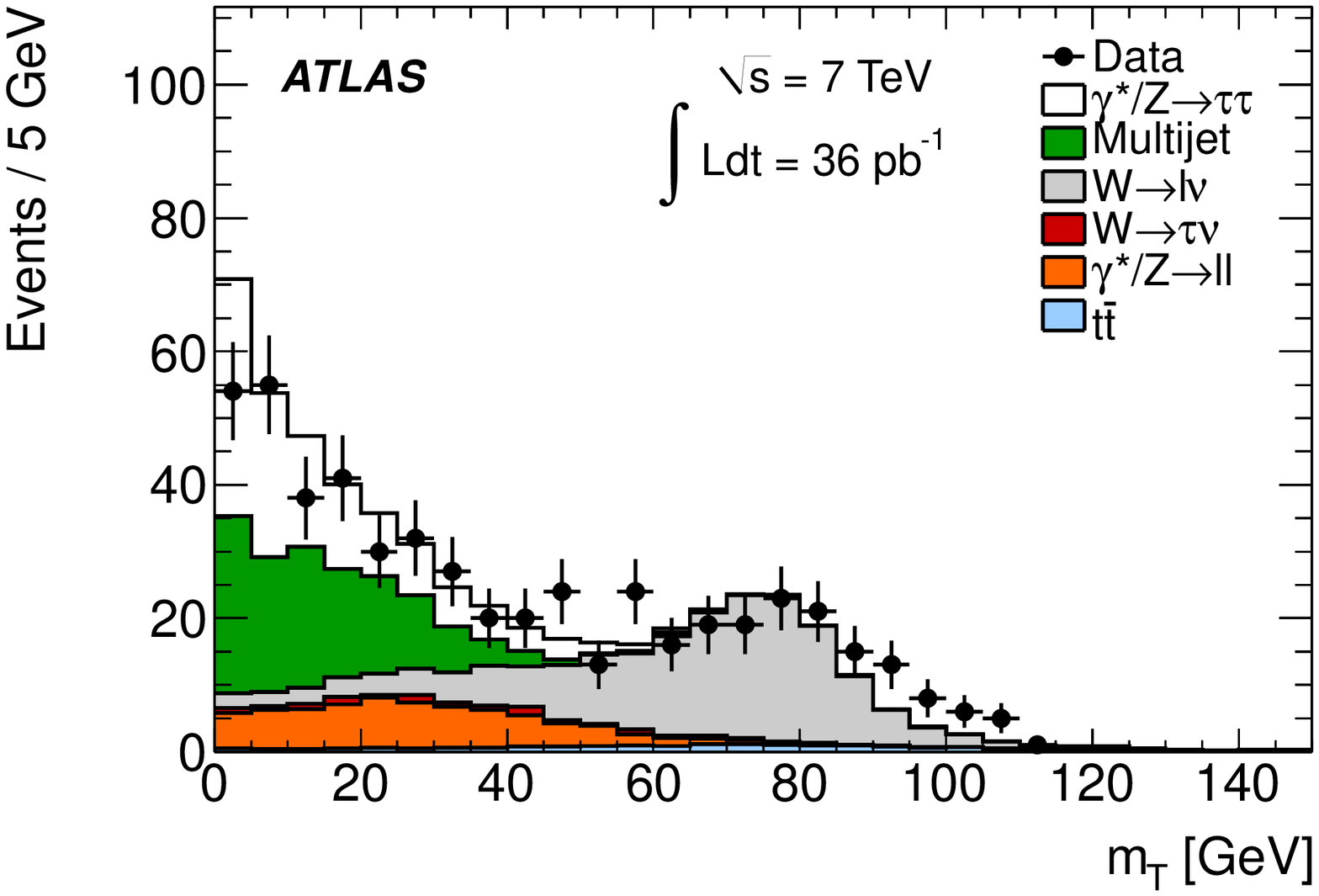}
        \label{fig:lephad_ele_h_trans_mass}
    }
    \caption{The distributions of $\sum\cos\Delta\phi$ are shown for the (a) $\tau_{\mu}\tau_{h}$  and (b) $\tau_{e}\tau_{h}$  final states.
        The distributions of the transverse mass, \mt, are shown for the (c) $\tau_{\mu}\tau_{h}$ and (d) $\tau_{e}\tau_{h}$ final states.  All distributions
        are shown after the object selection for the given final state and
	after requiring exactly one muon or electron candidate. 
        A requirement on the charge of the $\tau$ candidate to be of opposite sign to that of the lepton is also 
        applied. 
        The multijet background is estimated from data according to the method described in Section \ref{sec:background_estimation}; 
         all other processes are estimated using MC simulations.
         \label{fig:event_selection/wcuts}}
\end{figure*}

\subsubsection{$\tau_{\ell}\tau_{h}$ final states}
The multijet background is largely suppressed by the $\tau$ identification and lepton 
isolation requirements previously discussed. Events due to  \mbox{$W\rightarrow\ell\nu$}, \mbox{$W\rightarrow\tau\nu\rightarrow\ell\nu\nu\nu$},
and $\gamma^*/Z\rightarrow\ell\ell$ decays 
can be rejected with additional event-level selection criteria.

Any event with more than one muon or electron candidate is vetoed, which strongly suppresses background 
from \mbox{$\gamma^*/Z\rightarrow\ell\ell$ + jets} events. To increase the background rejection, the selection criteria for 
the second lepton are relaxed with respect to those described in Section~\ref{sec:rec_obj}:  the inner detector track quality 
requirements are dropped for the muons, while the electrons need 
only pass the ``medium''  selection and have $E_{T}>15~\GeV$.

In order to suppress the $W$+jets background, two additional selection criteria are applied.
For signal events the $\met$ vector is expected to fall in the azimuthal range spanned by the decay products, 
while in  \mbox{$W \rightarrow \ell\nu + \mathrm{jets}$} events it will tend to point outside 
of the angle between the jet faking the $\tau$ decay products and the lepton.
Hence the discriminating variable $\sum\cos\Delta\phi$  is defined as
\begin{equation}
  \begin{split}
  \sum\cos\Delta\phi& = \cos\big(\phi(\ell) - \phi(\MET)\big)\\
  &\quad+ \cos\big(\phi(\tau_\mathrm{h}) - \phi(\MET)\big)\:.
  \end{split}
\end{equation}
The variable \mbox{$\sum\cos\Delta\phi$} is positive when the $\met$ vector points towards
the direction bisecting the decay products and is negative when it points away. 
The distributions of $\sum\cos\Delta\phi$ are shown in Figure~\ref{fig:lephad_muon_h_sum_cos_dphi} and 
\ref{fig:lephad_ele_h_sum_cos_dphi} for the $\tau_{\mu}\tau_{h}$ and $\tau_{e}\tau_{h}$ final states, respectively.
The peak at zero corresponds to \mbox{$\gamma^*/Z\rightarrow\tau\tau$} events where the decay products 
are back-to-back in the transverse plane. 
The
\mbox{$W + \mathrm{jets}$} backgrounds accumulate at negative \mbox{$\sum\cos\Delta\phi$}
whereas the \mbox{$\gamma^*/Z\rightarrow\tau\tau$} distribution has an asymmetric tail extending
into positive \mbox{$\sum\cos\Delta\phi$} values, corresponding to events where the $Z$ boson has
higher $\pT$.  Events are therefore selected by requiring \mbox{$\sum\cos\Delta\phi > -0.15$}.
Even though the resolution of the $ \phi(\MET)$  direction is degraded for low values of $\MET$, 
this has no adverse effect on the impact of this selection, as such events correspond to \mbox{$\sum\cos\Delta\phi \sim 0$} 
and hence pass the selection.

To further suppress the \mbox{$W + \mathrm{jets}$} background, the transverse mass, defined as

\begin{equation}
    m_\mathrm{T}
     = \sqrt{2 \: p_\mathrm{T}(\ell) \cdot \MET \cdot \left( 1 - \cos\Delta\phi(\ell, \MET) \right)}\;,
\end{equation}
is required to be $m_\mathrm{T} < 50\GeV$.
Figures~\ref{fig:lephad_muon_h_trans_mass} and 
\ref{fig:lephad_ele_h_trans_mass} show the distribution of $m_T$ for the $\tau_{\mu}\tau_{h}$ and $\tau_{e}\tau_{h}$ final states, respectively.

\begin{figure*}
    \centering
    \subfigure[$\tau_{\mu}\tau_{h}$ final state]{
        \includegraphics[width=0.48\textwidth]{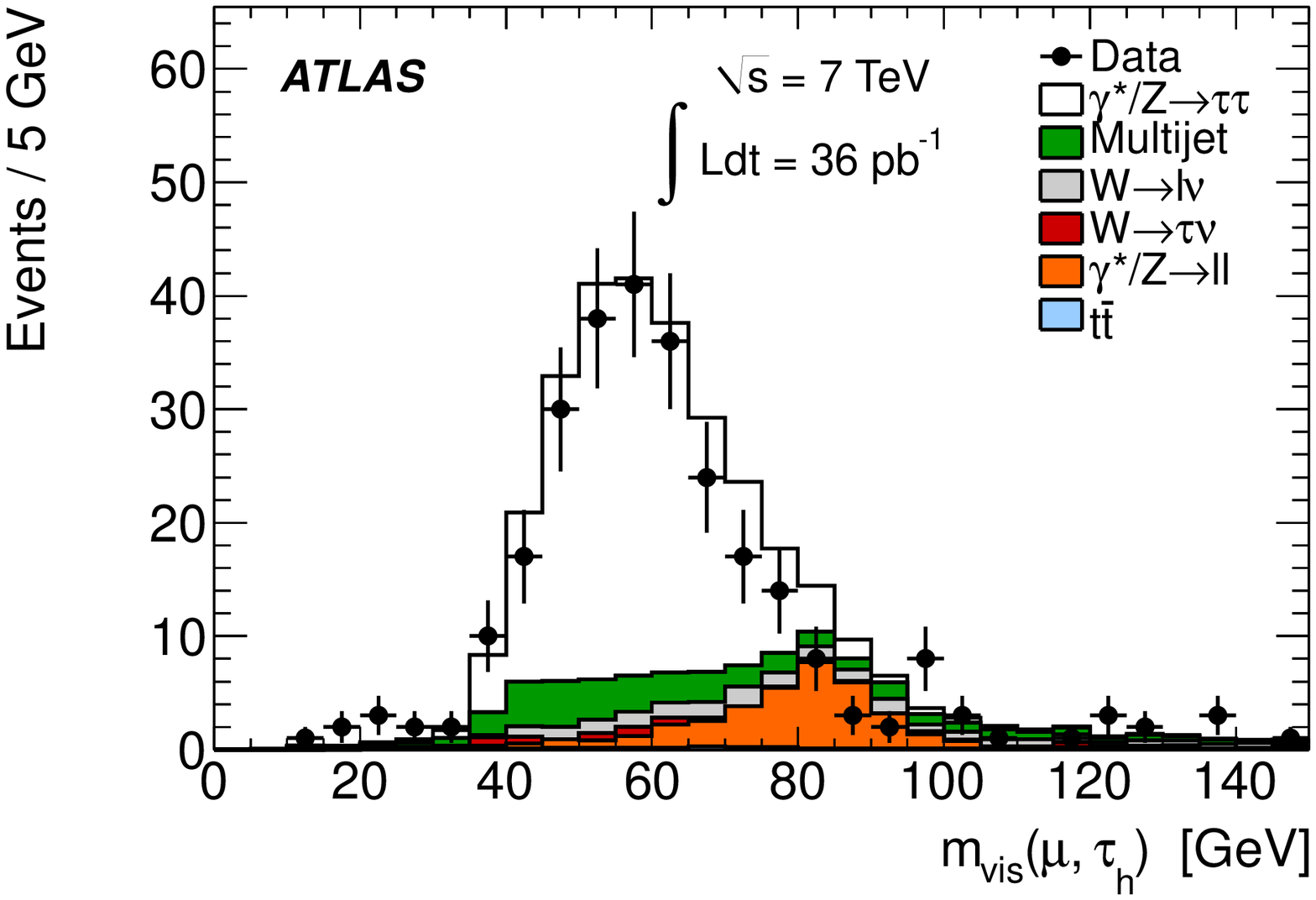}
        \label{fig:lephad_MuChannel_SigRegAllh_vis_mass}
    }   
    \subfigure[$\tau_{e}\tau_{h}$ final state]{
        \includegraphics[width=0.48\textwidth]{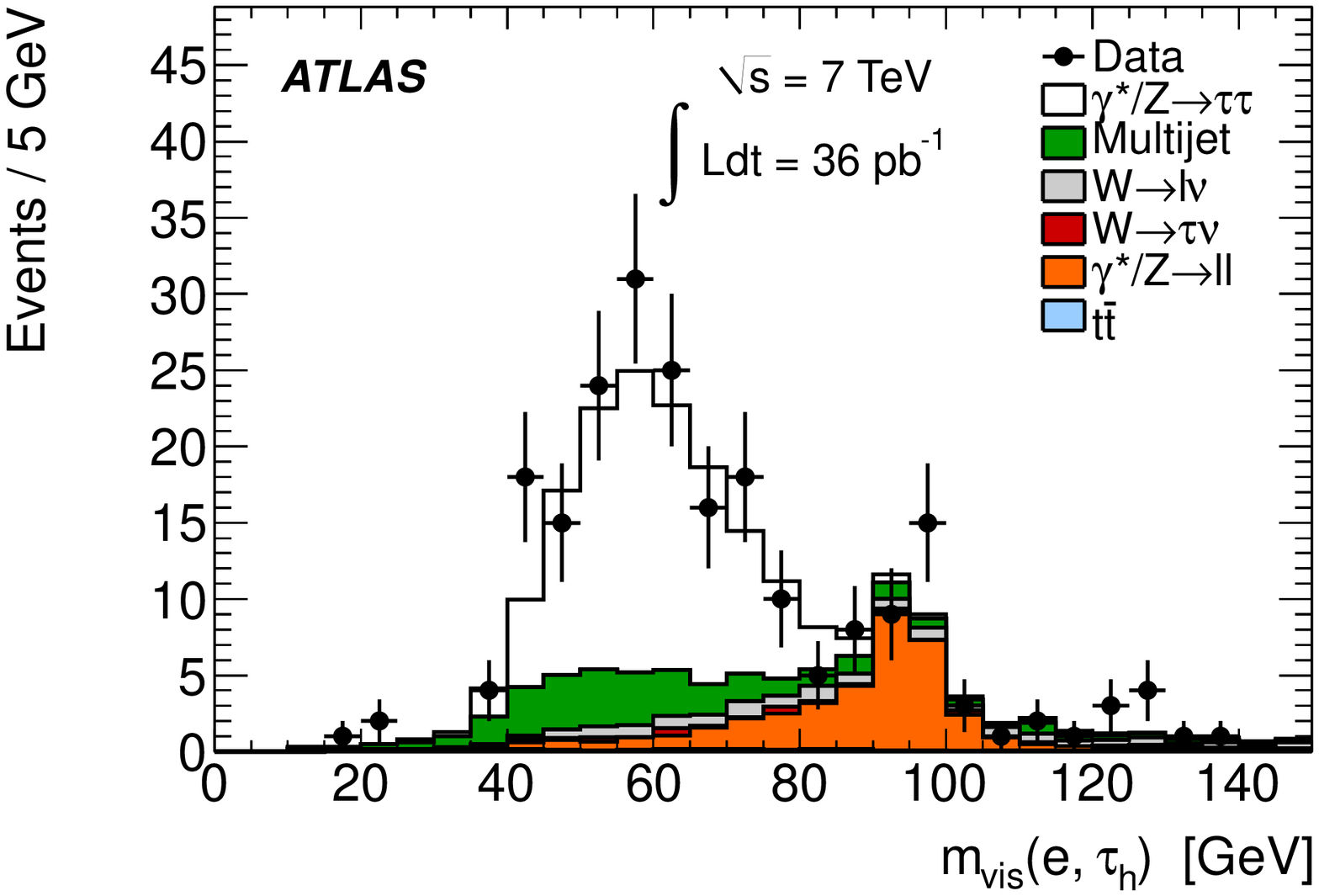}
        \label{fig:lephad_EChannel_SigRegAllh_vis_mass}
    }
    \caption{
        The distributions of the visible mass of the combination of the $\tau$ candidate and the
        lepton are shown for the (a) $\tau_{\mu}\tau_{h}$ and (b) $\tau_{e}\tau_{h}$  final states. 
        These distributions are shown after the full event selection, 
	except for the visible mass window requirement.
        \label{fig:event_selection/h_vis_mass}}
\end{figure*}

The visible mass $m_{\mathrm{vis}}$ is defined as the invariant mass of 
the visible decay products of the two $\tau$ leptons.
Selected events are required to have a visible mass in the range $35 < m_{\mathrm{vis}} < 75 \GeV$.
This window is chosen to include the bulk of the signal, while avoiding
background contamination from $Z\rightarrow\ell\ell$ decays. 
For $Z\rightarrow\mu\mu$ events the peak is at slightly lower values than for $Z\rightarrow ee$ events, for two reasons: 
muons misidentified as $\tau$ candidates leave less energy in the calorimeter compared to misidentified electrons, 
and the proportion of events where the $\tau$ candidate arises from a misidentified jet, as opposed to a misidentified lepton, is higher in $Z\rightarrow\mu\mu$ events.

Furthermore, the chosen $\tau$ candidate is required to have exactly 1 or 3 associated tracks and a
reconstructed charge of unit magnitude, characteristic of hadronic $\tau$ decays.
The charge is determined as the sum of the charges of the
associated tracks. Finally, the chosen $\tau$  candidate and the chosen lepton are required to have opposite charges  
as expected from  \mbox{$Z\rightarrow\tau\tau$} decays.

\begin{figure*}
    \centering
    \subfigure[$\tau_{\mu}\tau_{h}$ final state]{
        \includegraphics[width=0.48\textwidth]{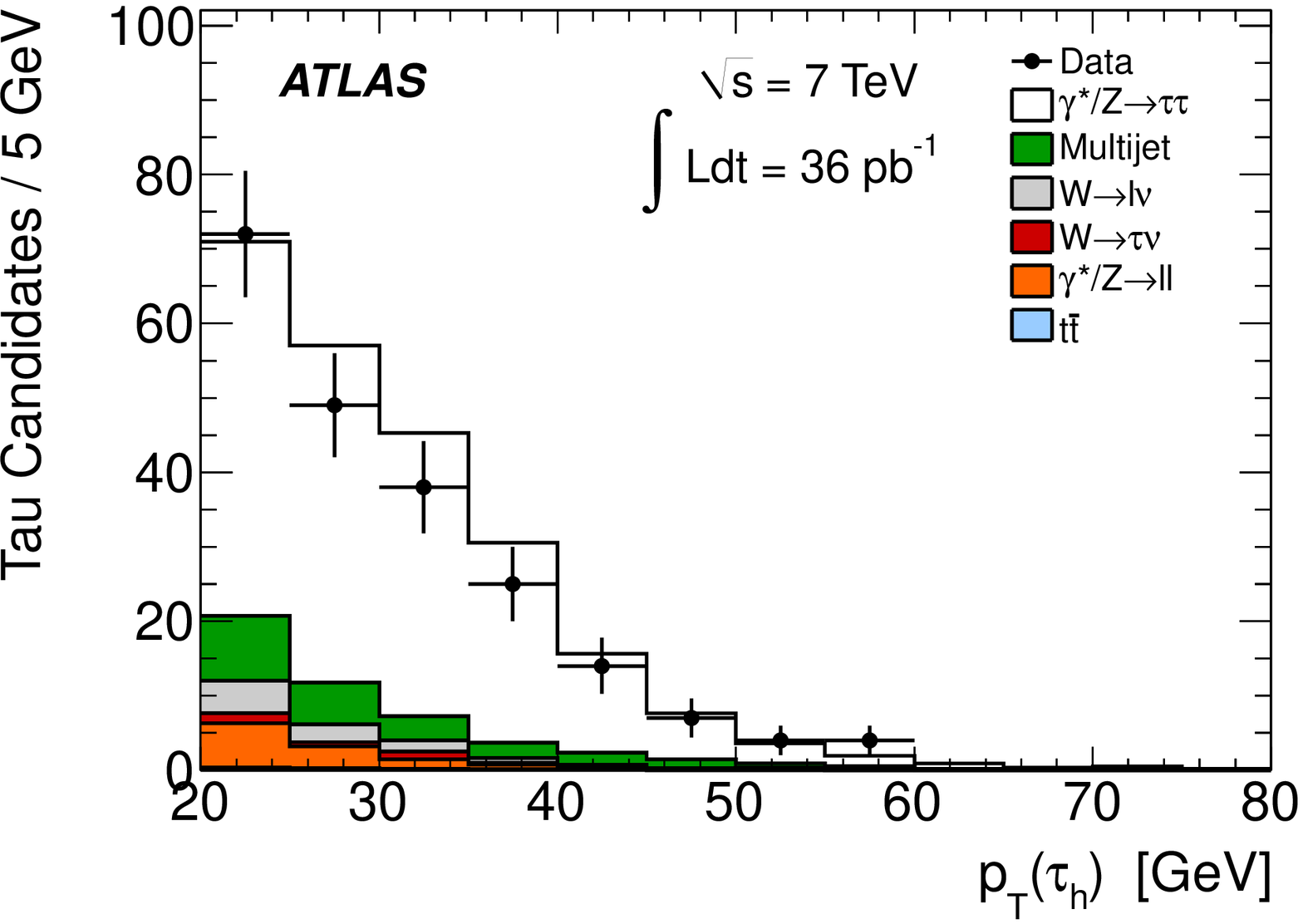}
        \label{fig:lephad_MuChannel_SigRegAllh_tau_pt}
    }   
    \subfigure[$\tau_{e}\tau_{h}$ final state]{
        \includegraphics[width=0.48\textwidth]{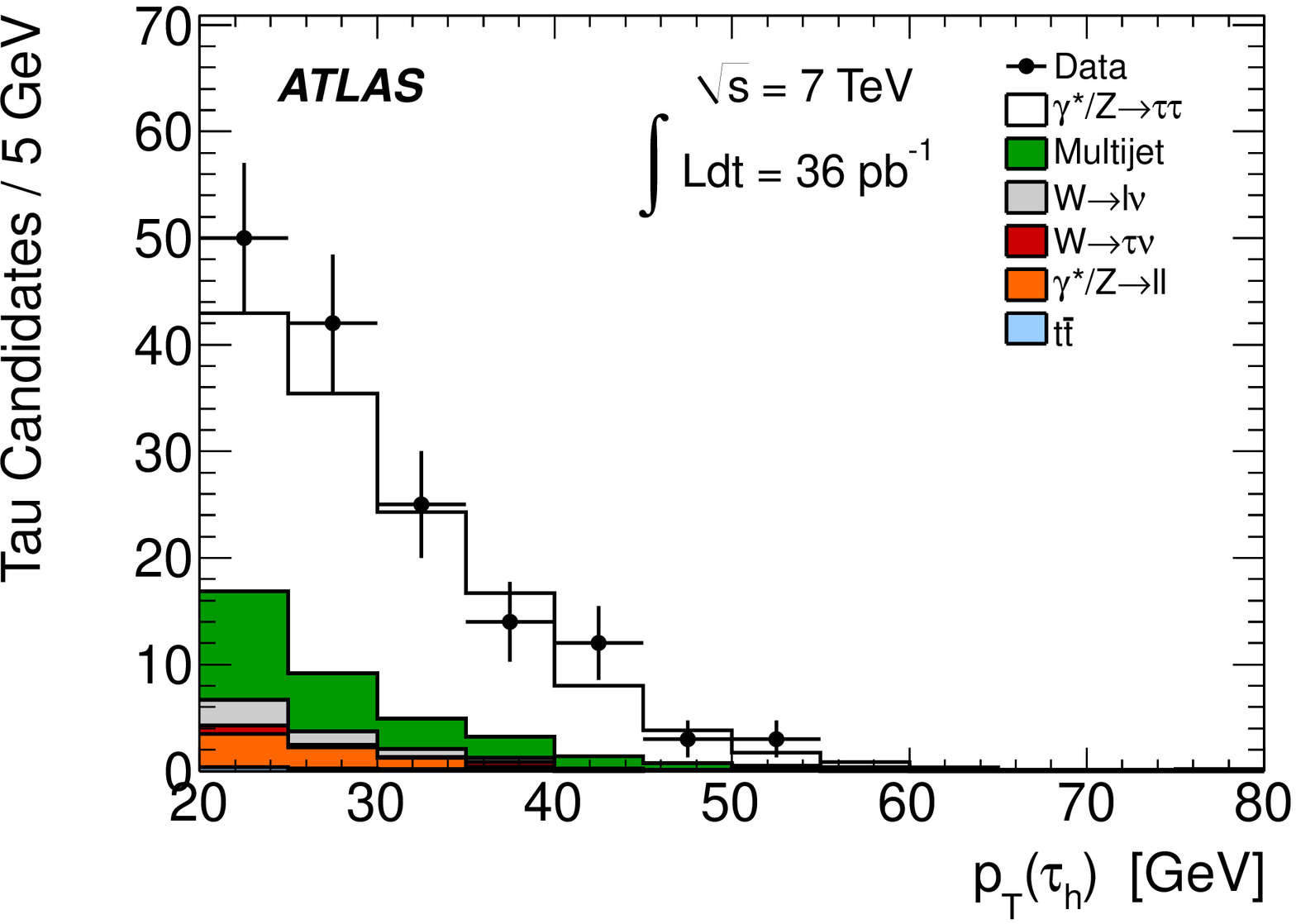}
        \label{fig:lephad_EChannel_SigRegAllh_tau_pt}
    } \\*
    \subfigure[$\tau_{\mu}\tau_{h}$ final state]{
        \includegraphics[width=0.48\textwidth]{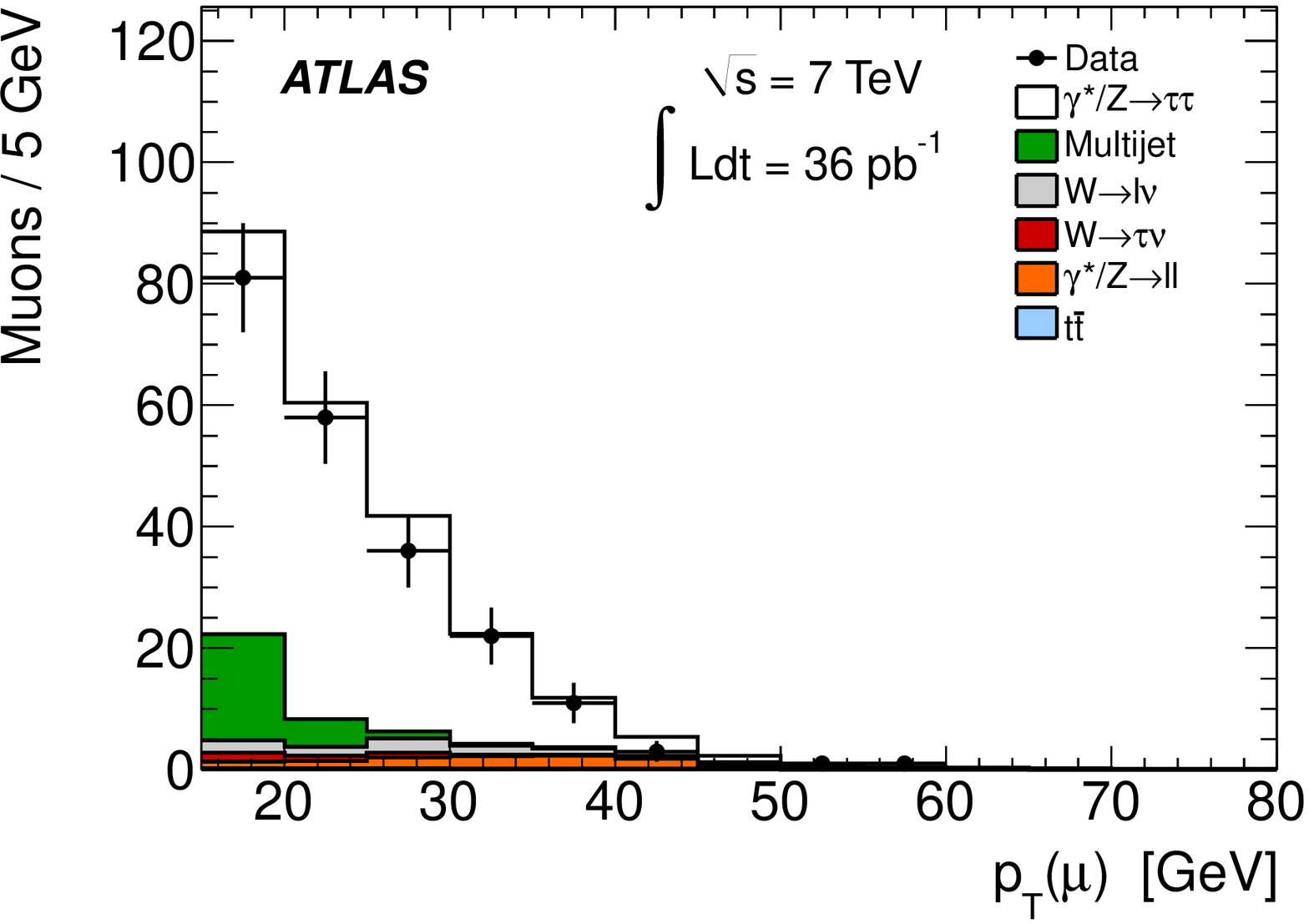}
        \label{fig:lephad_MuChannel_SigRegAllh_mu_pt}
    }   
    \subfigure[$\tau_{e}\tau_{h}$ final state]{
        \includegraphics[width=0.48\textwidth]{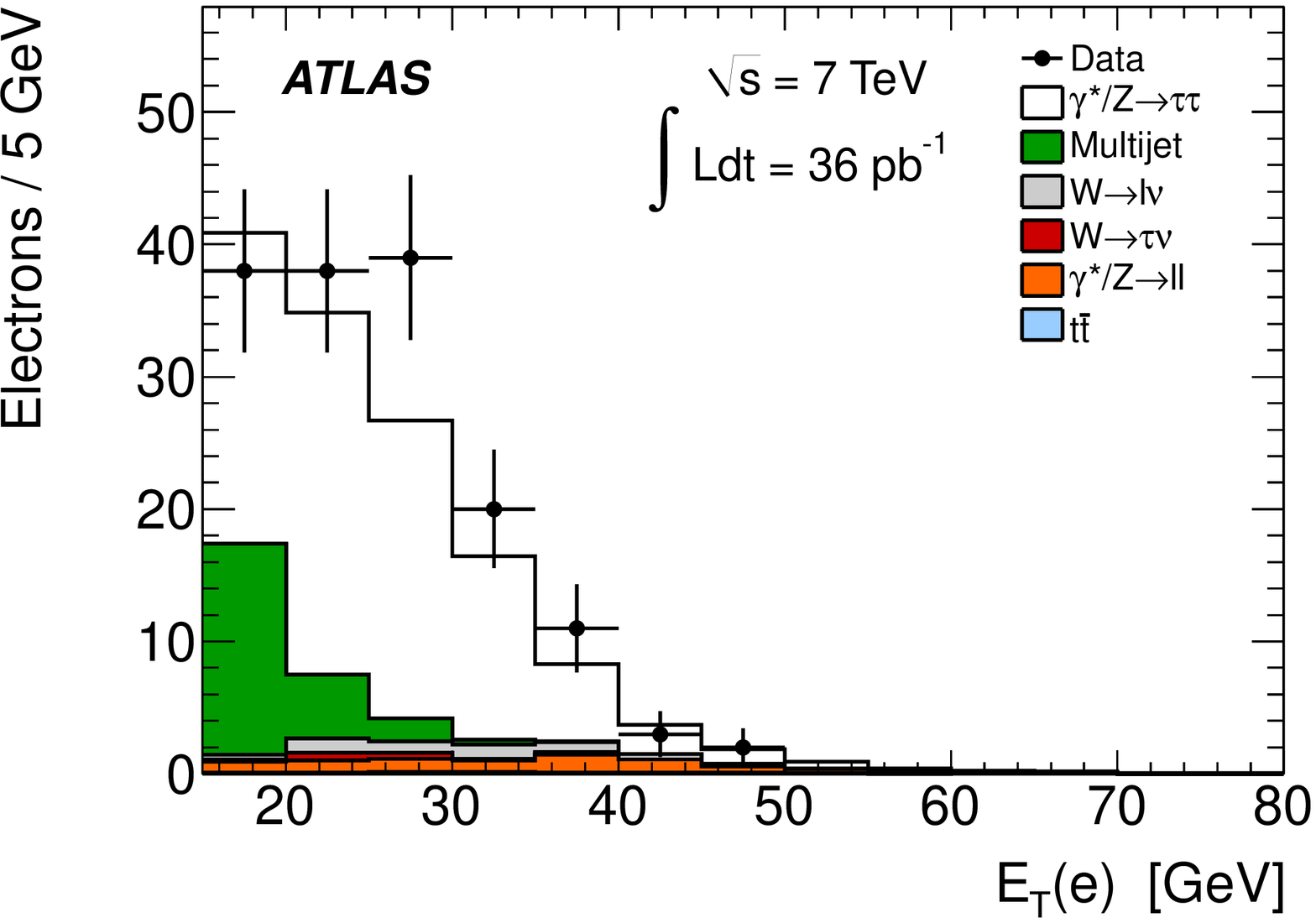}
        \label{fig:lephad_EChannel_SigRegAllh_el_pt}
    }
    \caption{
        Distributions of the $p_\mathrm{T}$ of the $\tau$ candidate and of the
	muon and $E_\mathrm{T}$ of the electron, for events passing all signal
	selections
        for the $\tau_{\mu}\tau_{h}$ and $\tau_{e}\tau_{h}$ final states.
	\label{fig:observation/tau_kin}}
\end{figure*}

\begin{figure*}
    \centering
    \subfigure[$\tau_{\mu}\tau_{h}$ final state]{
        \includegraphics[width=0.48\textwidth]{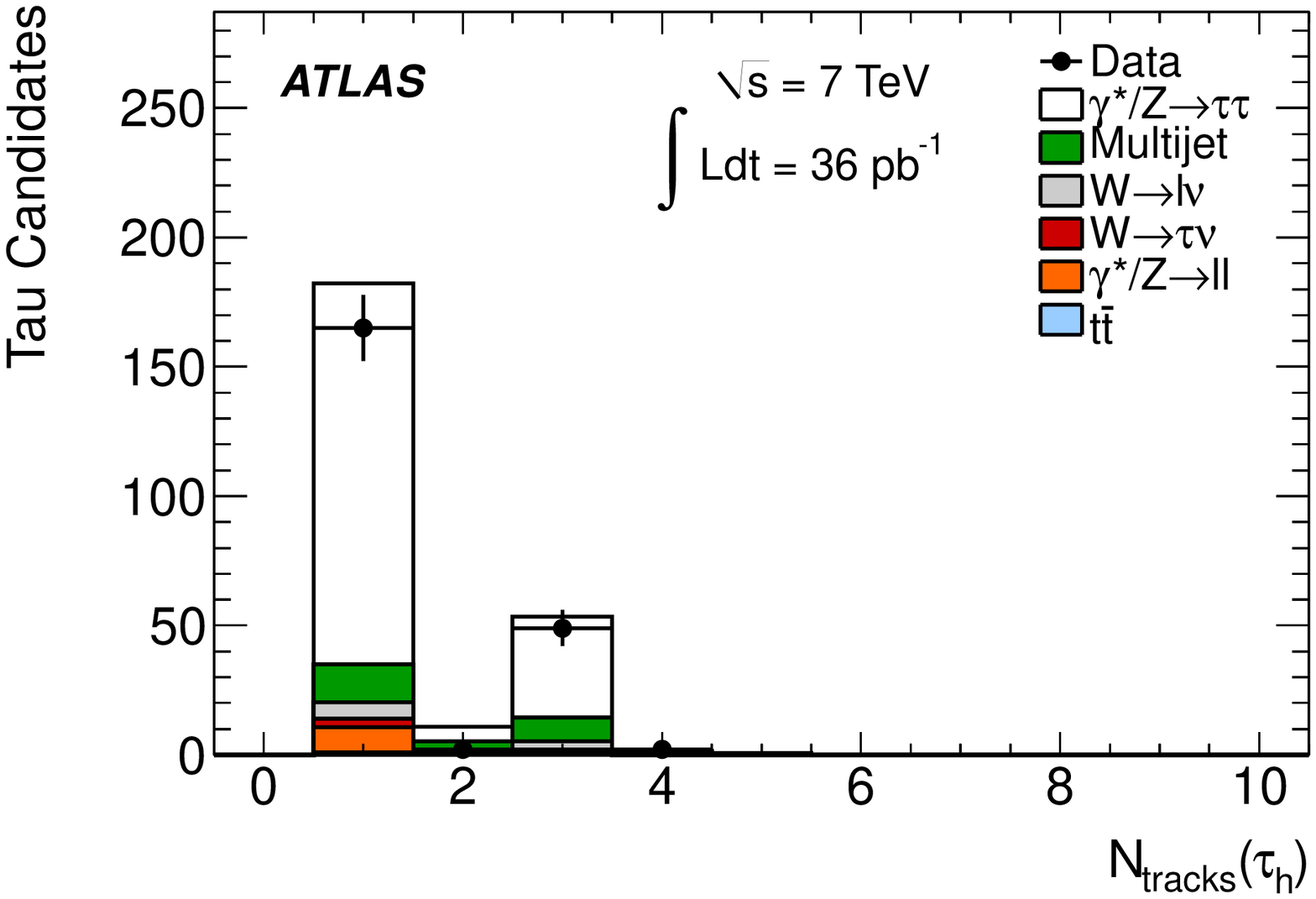}
        \label{fig:lephad_mu_h_tau_numTrack}
    }   
    \subfigure[$\tau_{e}\tau_{h}$ final state]{
        \includegraphics[width=0.48\textwidth]{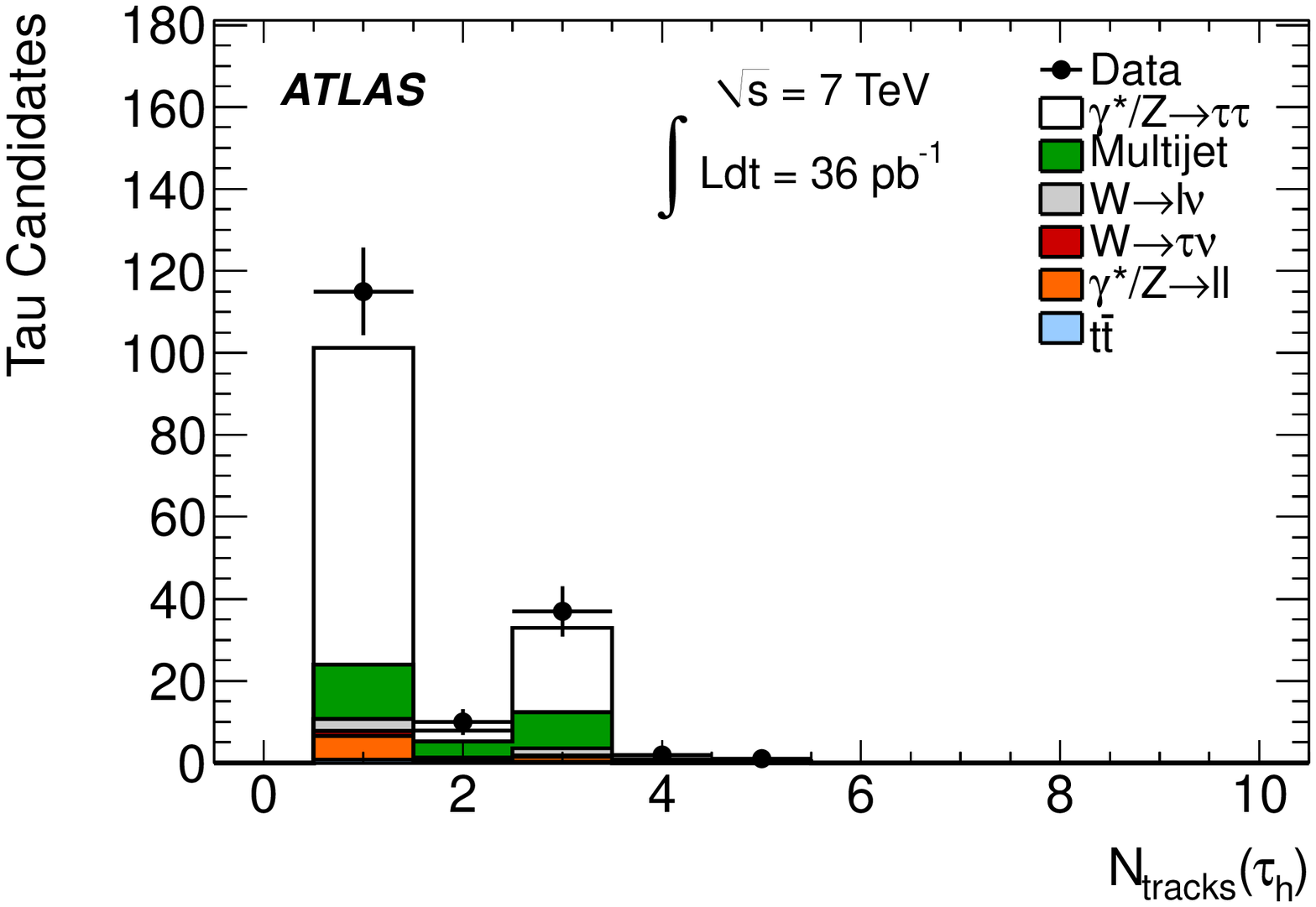}
        \label{fig:lephad_ele_h_tau_numTrack}
    } \\*
    \caption{
Distribution of the number of tracks associated to $\tau$ candidates 
after the full selection, including the opposite-charge requirement  
for the $\tau$ candidate and the lepton, except the requirement on the number of tracks and on the magnitude 
      of the $\tau$ charge.
      \label{fig:observation/numTrack}}
\end{figure*}

The distribution of the visible mass after the full selection except the visible mass window requirement is 
shown in Figure~\ref{fig:event_selection/h_vis_mass}. The distributions of the lepton and $\tau$ candidate $\pt$, for events passing all signal selection criteria, are shown in Figure~\ref{fig:observation/tau_kin}.
The  $\tau$ candidate track distribution after the full selection 
except the  requirements on the number of associated tracks and on the magnitude of the $\tau$ charge is shown 
in Figure~\ref{fig:observation/numTrack}.

\subsubsection{$\tau_{e}\tau_{\mu}$ final state}
The $\tau_e \tau_\mu$ events are characterized by the presence of two oppositely charged and isolated 
leptons in the final state. 
Thus exactly one electron and one muon candidate of opposite electric charge,
which pass the selections described in Section~\ref{sec:rec_obj}, are required.
For events that contain two leptons of different flavors, the contributions 
from $\gamma^{*}/Z \ra ee$ and $\gamma^{*}/Z \ra \mu\mu$ processes is small.
The remaining background is therefore due to $W$ and $Z$ leptonic decays, where an additional
real or fake lepton comes from jet fragmentation.

\begin{figure}
    \centering
    \subfigure[$\tau_{e}\tau_{\mu}$  final state]{
        \includegraphics[width=0.48\textwidth]{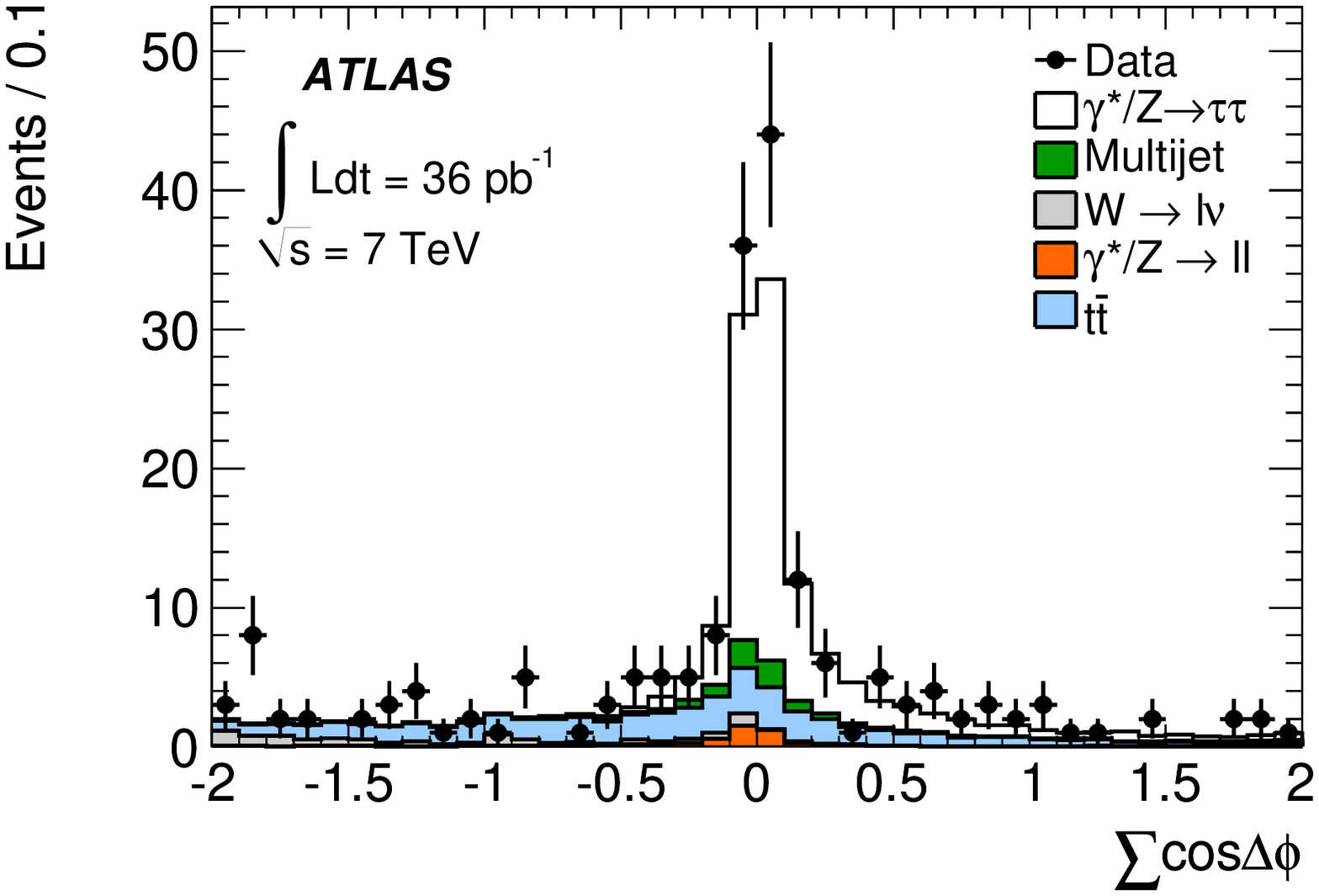}
        \label{fig:muele_IsoOSSumCosDPhiLepMET}
    } 
    \subfigure[$\tau_{e}\tau_{\mu}$  final state]{
        \includegraphics[width=0.48\textwidth]{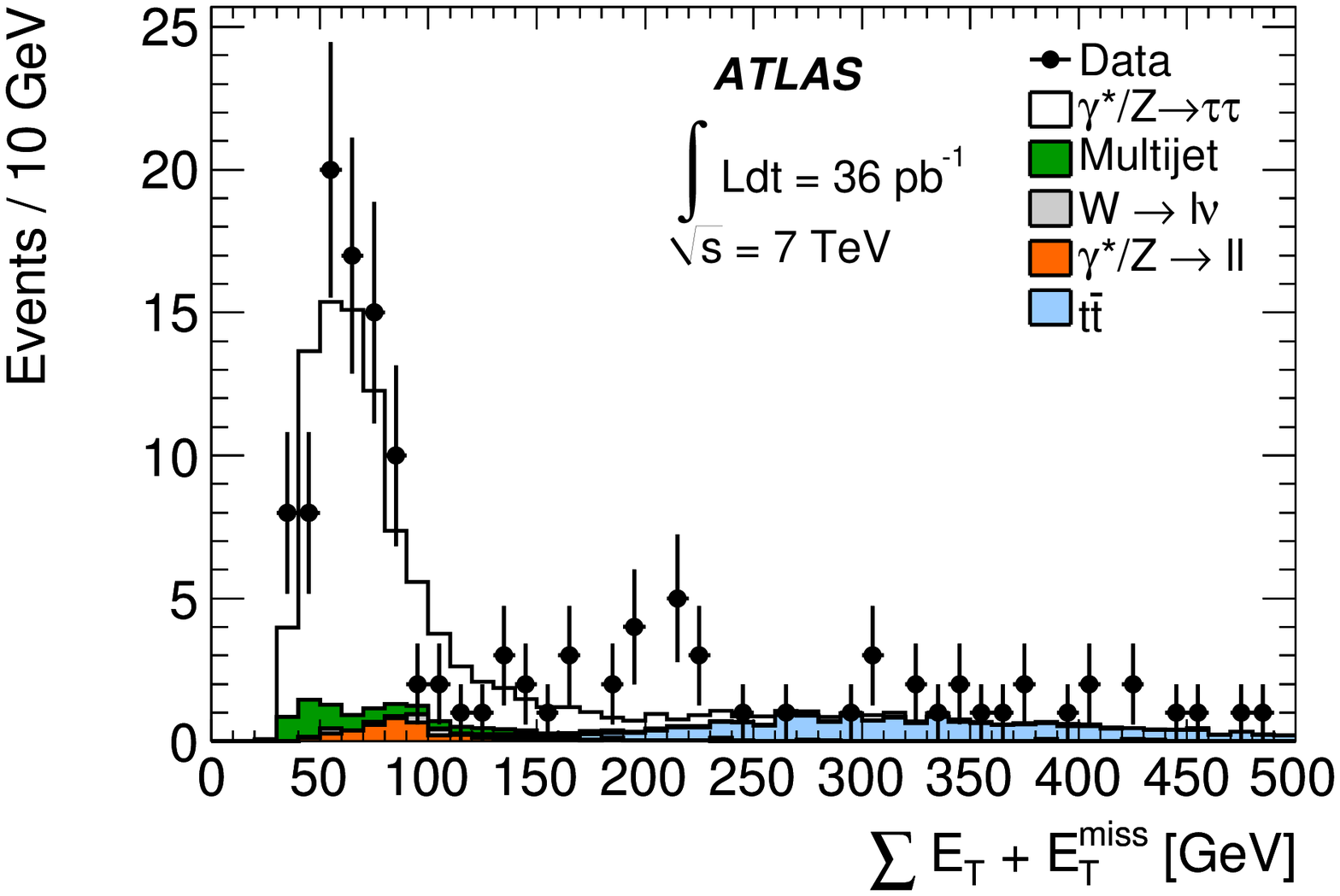}
        \label{fig:muele_IsoOSSumEtMET}
    }    
    \caption{Distributions of the variables (a) $\sum \mathrm{cos}\Delta\phi$, after 
     the lepton isolation selection, and (b) $\sumetmet$  after the $\sum \mathrm{cos}\Delta\phi$ selection,
     for the $\tau_{e}\tau_{\mu}$  final state. 
     The multijet background is estimated from data according to the method described in Section \ref{sec:background_estimation}; 
     all other processes are estimated using MC simulations.
     \label{fig:emu_sumcosmet}}
\end{figure}

To reduce the \Wen, \Wmn, and \ttbar~ backgrounds, the requirement~\mbox{$\sum\cos\Delta\phi > -0.15$} 
is applied as in the semileptonic final states.
Figure \ref{fig:emu_sumcosmet} shows the distribution of $\sum\cos\Delta\phi$ after 
the previous selection criteria.

A further requirement is made to reduce the \ttbar~ background.
Unlike for the signal, the topology of \ttbar~ events is characterized by the presence
of high-$\pt$ jets and leptons, as well as large \met~.

Hence the variable
\begin{equation}
        \sumetmet = \et(e) + \pt(\mu) + \sum_{jets} \pt + \MET
\end{equation}
is defined, where the electron and muon candidates, the jets and \met~ pass the selections described in Section~\ref{sec:rec_obj}.
The distribution of this variable for data and Monte Carlo after
the $\sum\cos\Delta\phi$ requirement is shown in Figure~\ref{fig:emu_sumcosmet}.
Requiring $\sumetmet< 150 \GeV$ rejects most of the \ttbar~ background.
 
\begin{figure*}
    \centering
    \subfigure[$\tau_{e}\tau_{\mu}$ final state]{
      \includegraphics[width=0.48\textwidth]{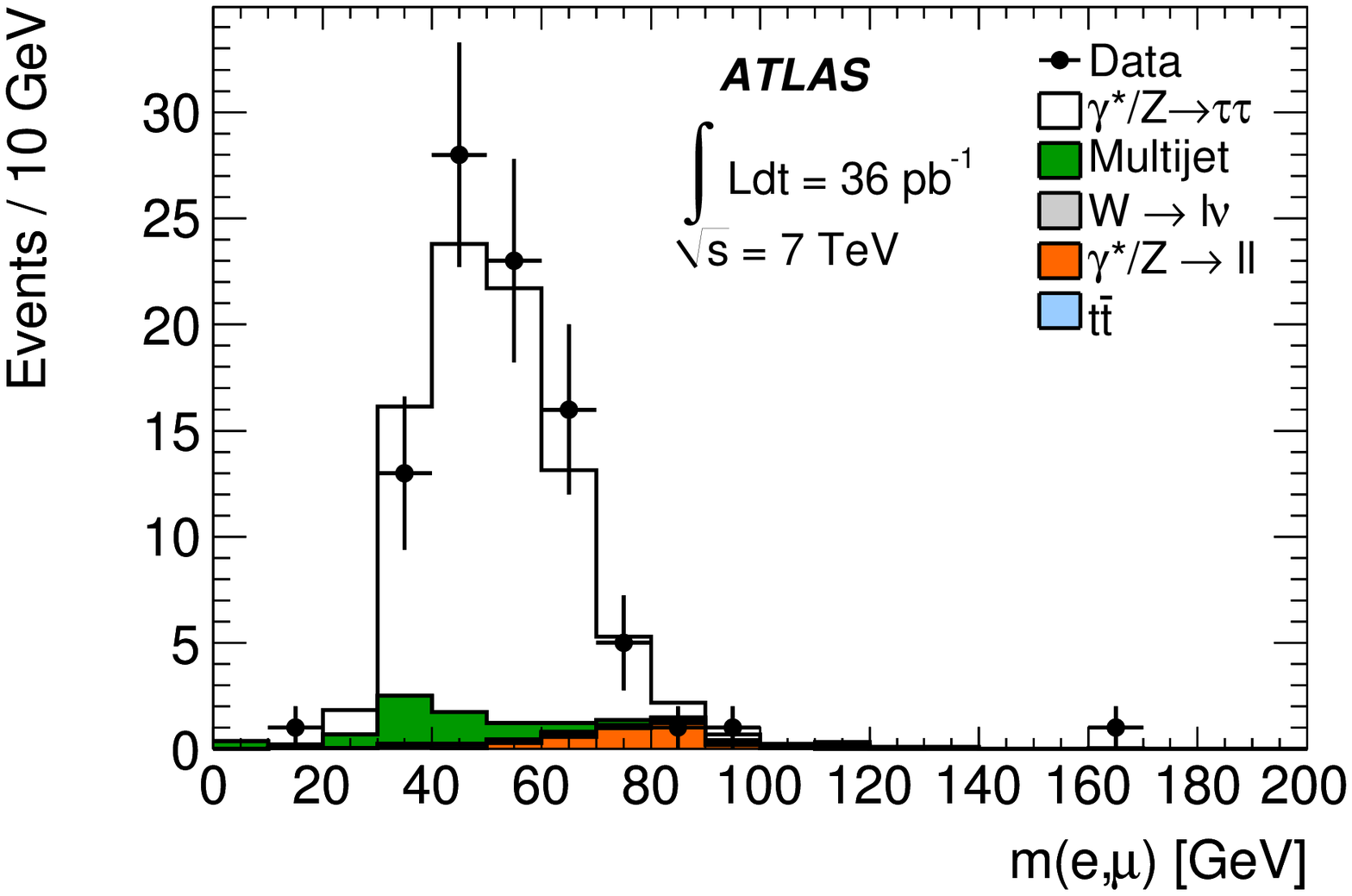}
      \label{fig:muele_vis_mass}
    }   
    \subfigure[$\tau_{\mu}\tau_{\mu}$ final state]{
      \includegraphics[width=0.48\textwidth]{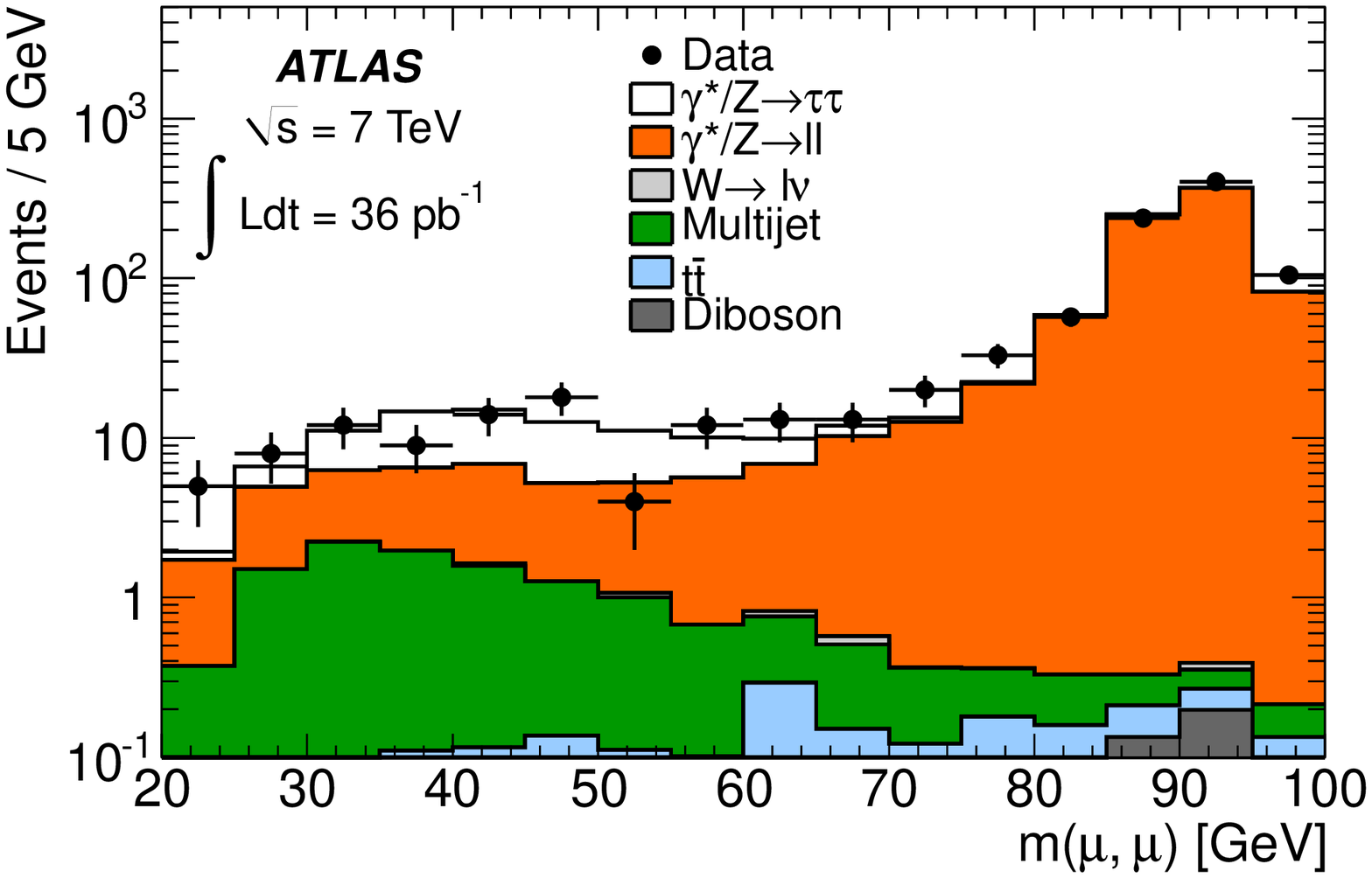}
      \label{fig:mumu_vis_mass}
    }
    \caption{
        The distributions of the visible mass for the (a) $\tau_{e}\tau_{\mu}$ and 
        (b) $\tau_{\mu}\tau_{\mu}$ final states, after all selections except the selection on the visible mass.
        \label{fig:leplep_vis_mass}}
\end{figure*}

\begin{figure}
    \centering
    \subfigure[$\tau_{e}\tau_{\mu}$ final state]{
        \includegraphics[width=0.48\textwidth]{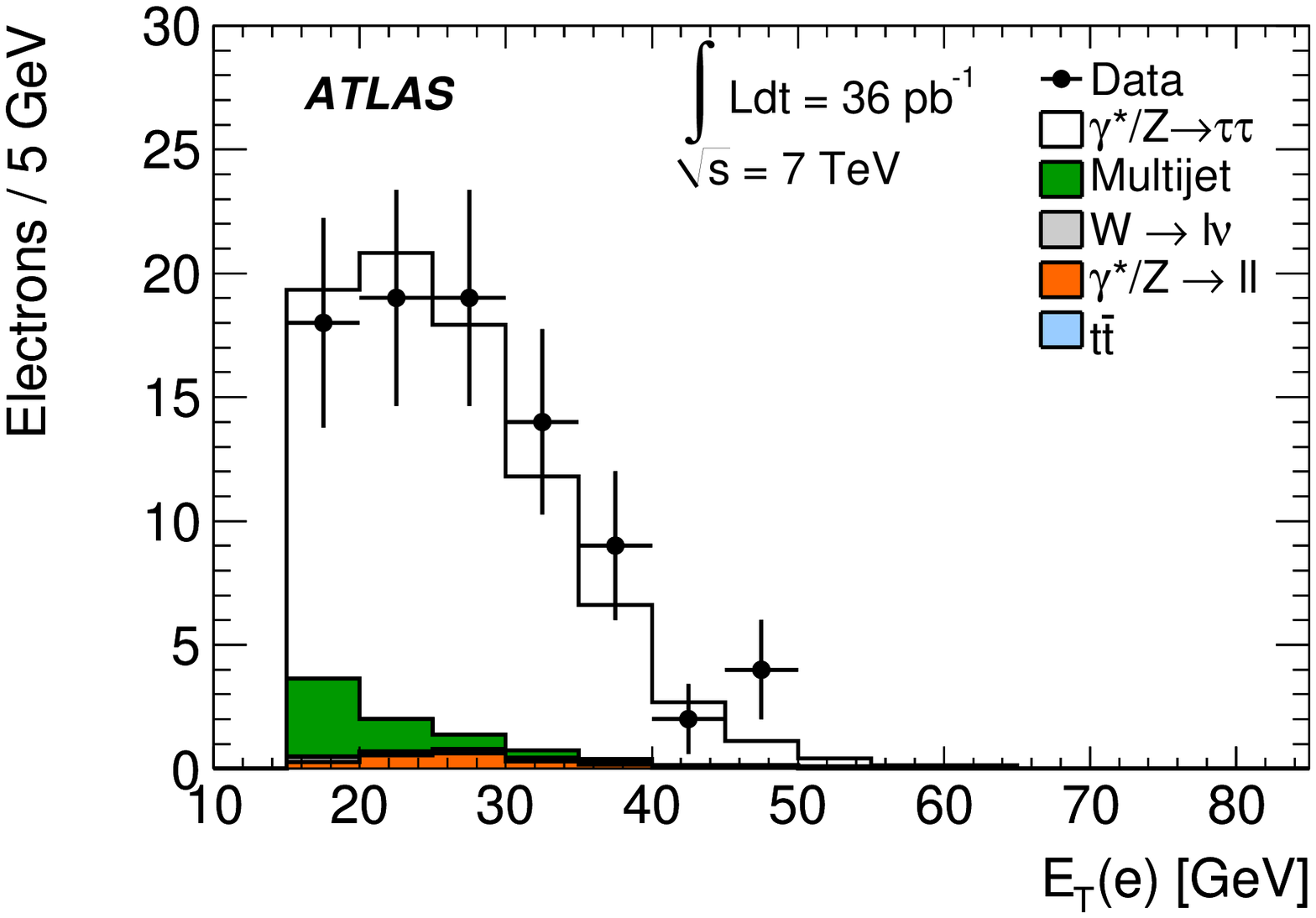}
        \label{fig:muele_ptele}
    }   
    \subfigure[$\tau_{e}\tau_{\mu}$ final state]{
       \includegraphics[width=0.48\textwidth]{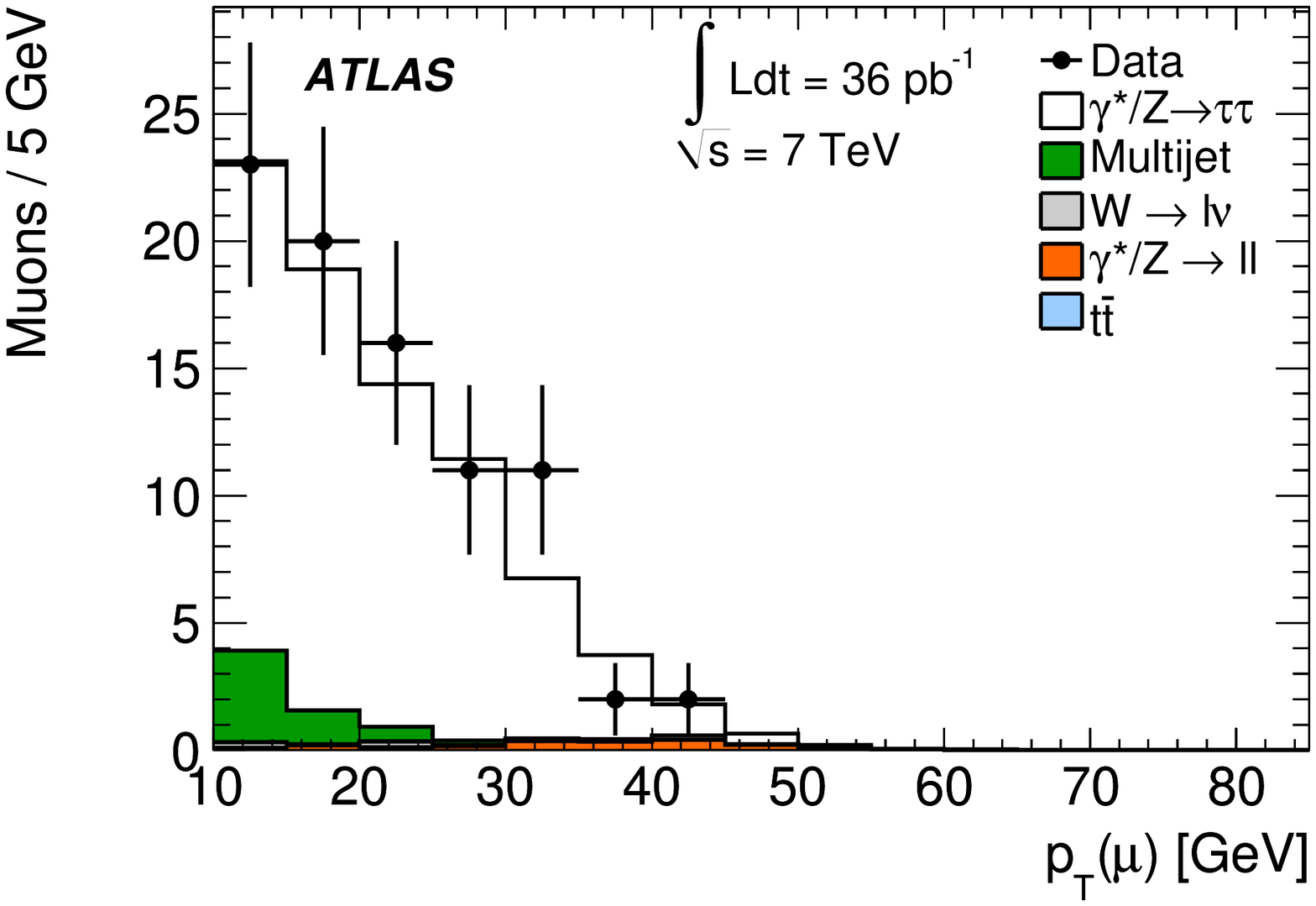}
        \label{fig:muele_ptmuon}
    }
    \caption{
        Distributions of the (a) $E_\mathrm{T}$ of the electron and (b) $p_\mathrm{T}$ of the muon, for events passing all selections
        for the $\tau_{e}\tau_{\mu}$ final state.
	\label{fig:muele/lep_kin}}
\end{figure}

Finally, since $\gamma^{*}/Z \rightarrow \ell \ell$ events are a small background in this final state,
the dilepton invariant mass is required to be within a wider range than in the semileptonic case: $25 < m_{e\mu} < 80 \gev$. 
Figure~\ref{fig:muele_vis_mass} shows the distribution of the visible mass.
Figure~\ref{fig:muele/lep_kin} shows the  $p_\mathrm{T}$ distributions of both leptons 
for events passing the full signal selection. 

\subsubsection{$\tau_{\mu}\tau_{\mu}$ final state}

The $\tau_\mu \tau_\mu$ final state is  characterized by two oppositely charged muons. 
Therefore only events that contain exactly two muon candidates with opposite charge that pass 
the selection criteria described in Section~\ref{sec:rec_obj} are considered, with the additional 
requirement that the leading muon has a transverse momentum greater than 15~\GeV.
The signal region for this final state is defined by the two muon candidates having an invariant mass  of 
$25 < m_{\mu\mu} < 65$~GeV. 

A boosted decision tree (BDT)~\cite{tmva} is used
to maximize the final signal efficiency and the  discrimination power against the background. 
The BDT is trained using  $Z\rightarrow \tau\tau$ 
Monte Carlo samples as  signal and $\gamma^*/ Z\rightarrow \mu\mu$ Monte Carlo samples as  background. No other backgrounds
are introduced in the training, in order to achieve the maximum separation between the signal and
the main ($\gamma^*/ Z \rightarrow \mu\mu$) background.
The BDT is trained after the selection of two oppositely charged muon candidates whose invariant mass fall within the signal region. 
To maximize the available Monte Carlo statistics for training and testing, no isolation requirements are applied to the muon candidates.

 \begin{figure*}
    \centering
    \subfigure[]{
        \includegraphics[width=0.48\textwidth]{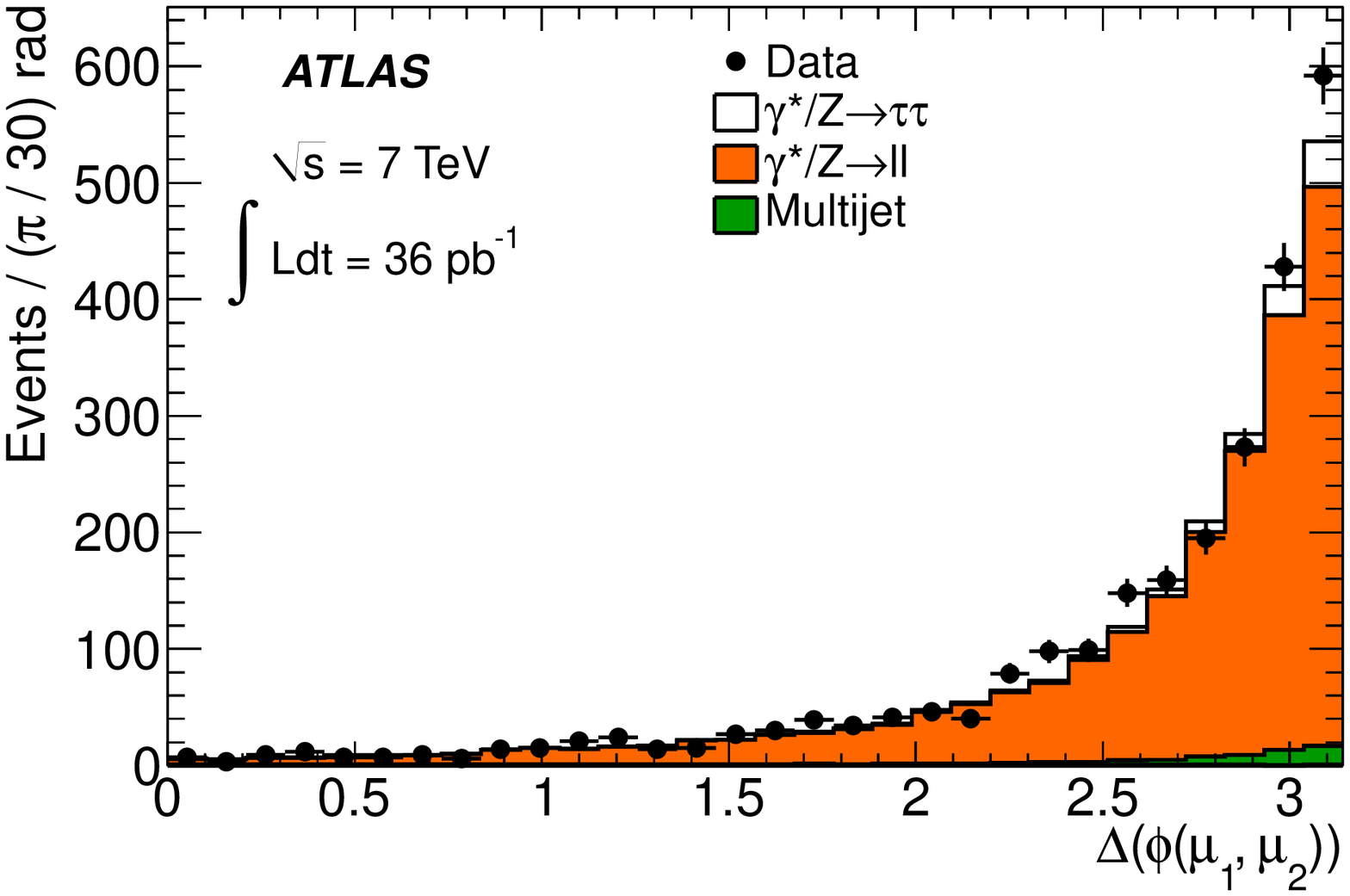}
        \label{fig:DeltaPhiLeptons}
    }   
    \subfigure[]{
        \includegraphics[width=0.48\textwidth]{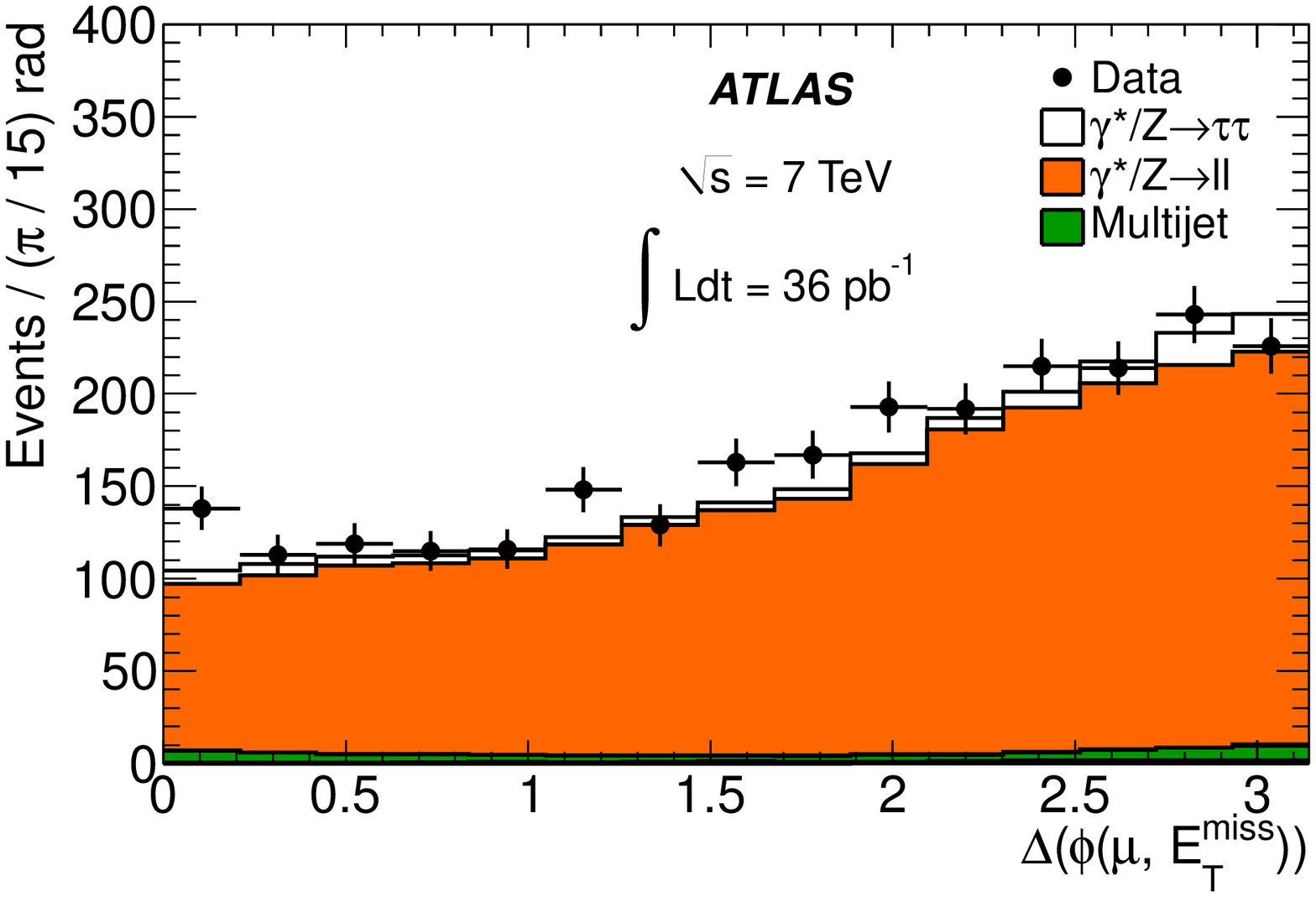}
        \label{fig:DeltaPhiLeptonMet} 
    }\\*  
    \subfigure[]{
        \includegraphics[width=0.48\textwidth]{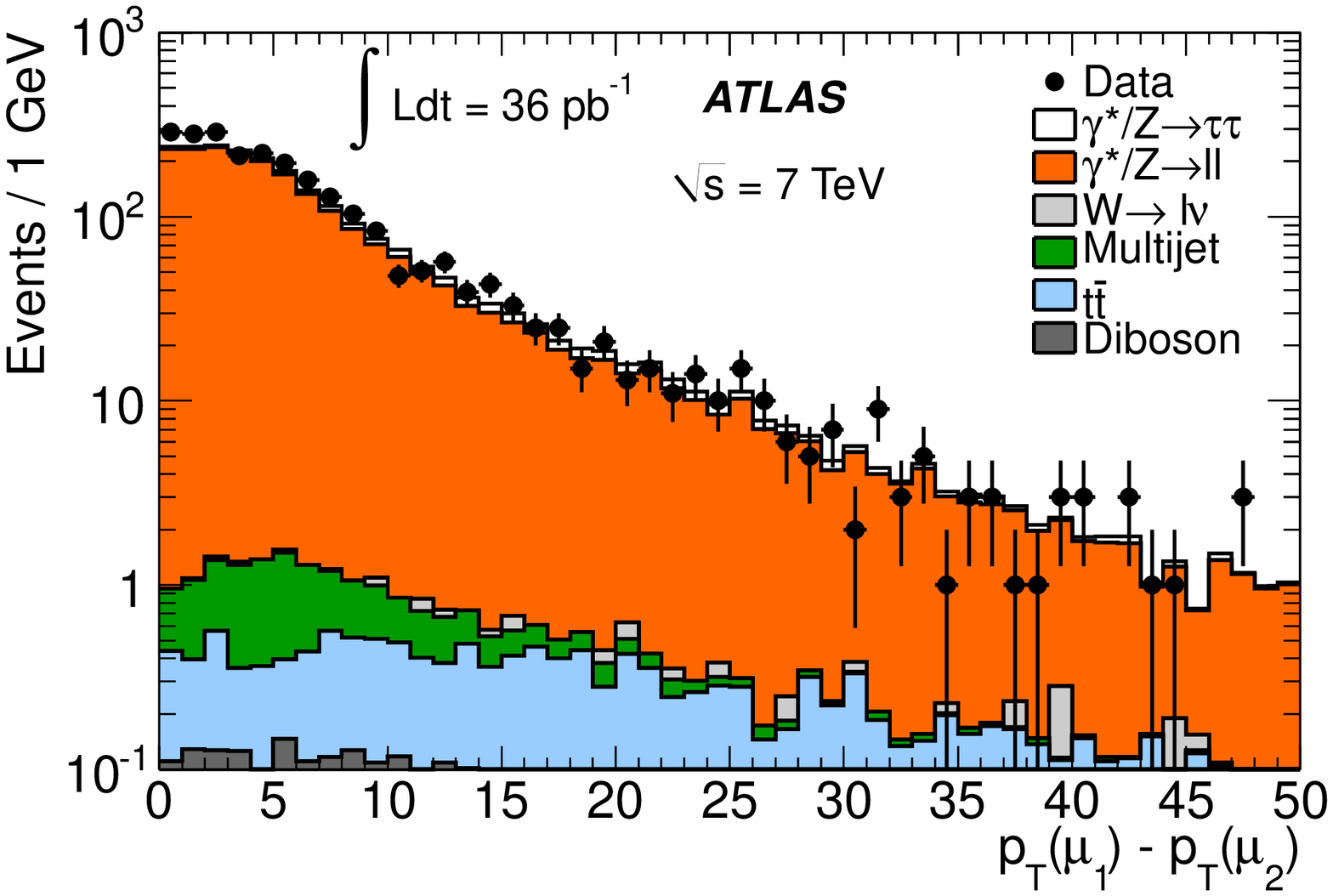}
        \label{fig:LeptonPtDifference}
    }   
    \subfigure[]{
        \includegraphics[width=0.48\textwidth]{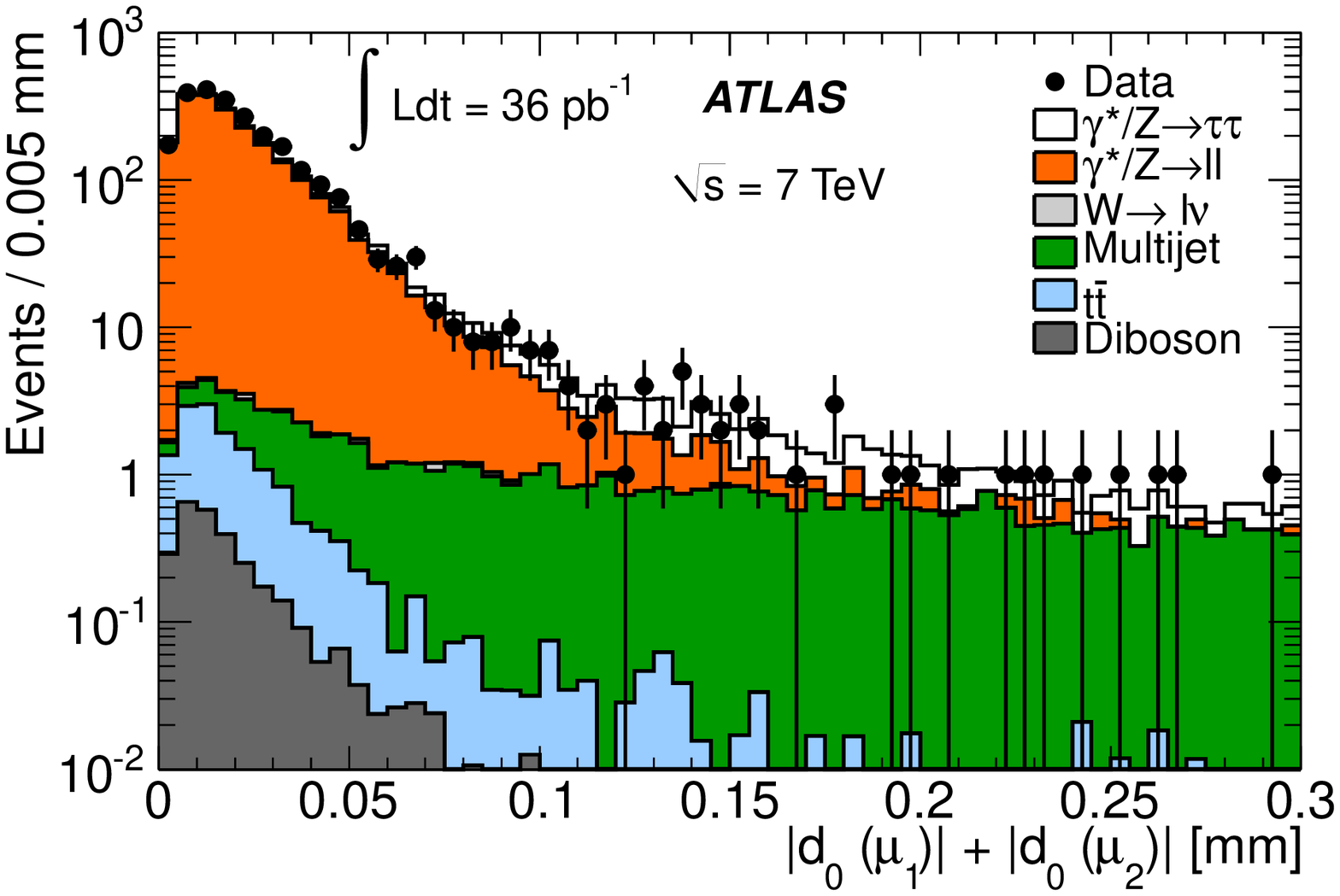}
        \label{fig:MuonD0} 
    }\\* 
 
    \caption{Distributions of some of the input variables to the BDT used to optimize the selection of the $\tau_{\mu}\tau_{\mu}$ final
     state. The data are compared to signal ($\gamma^*/ Z \rightarrow \tau\tau$) and $\gamma^*/ Z \rightarrow \mu\mu$ Monte Carlo samples. 
	The multijet background is estimated from data 
(see Section \ref{sec:background_estimation}).
The observed differences are
consistent with the estimated systematic uncertainties on the $\gamma^*/Z$ background normalization.
   \label{fig:bdt_var}}
\end{figure*}
 
\begin{figure}
    \centering
        \includegraphics[width=0.48\textwidth]{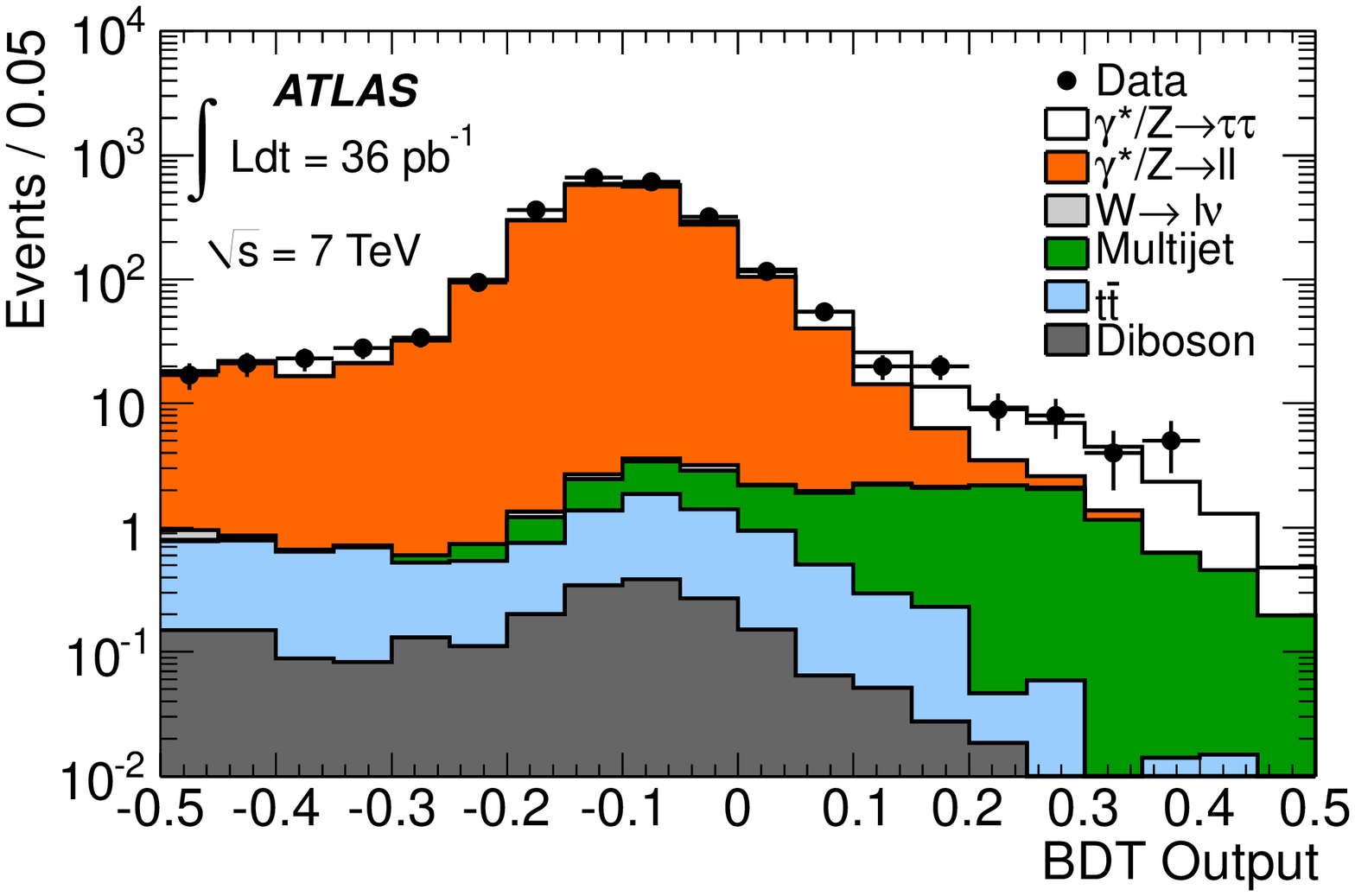}
        \caption{The distribution of the $\tau_{\mu}\tau_{\mu}$ BDT output for the data and the expected backgrounds. Events with a BDT output greater than 0.07 are selected. }\label{fig:BDTscore}
\end{figure}

\begin{figure}
    \centering
    \subfigure[]{
        \includegraphics[width=0.48\textwidth]{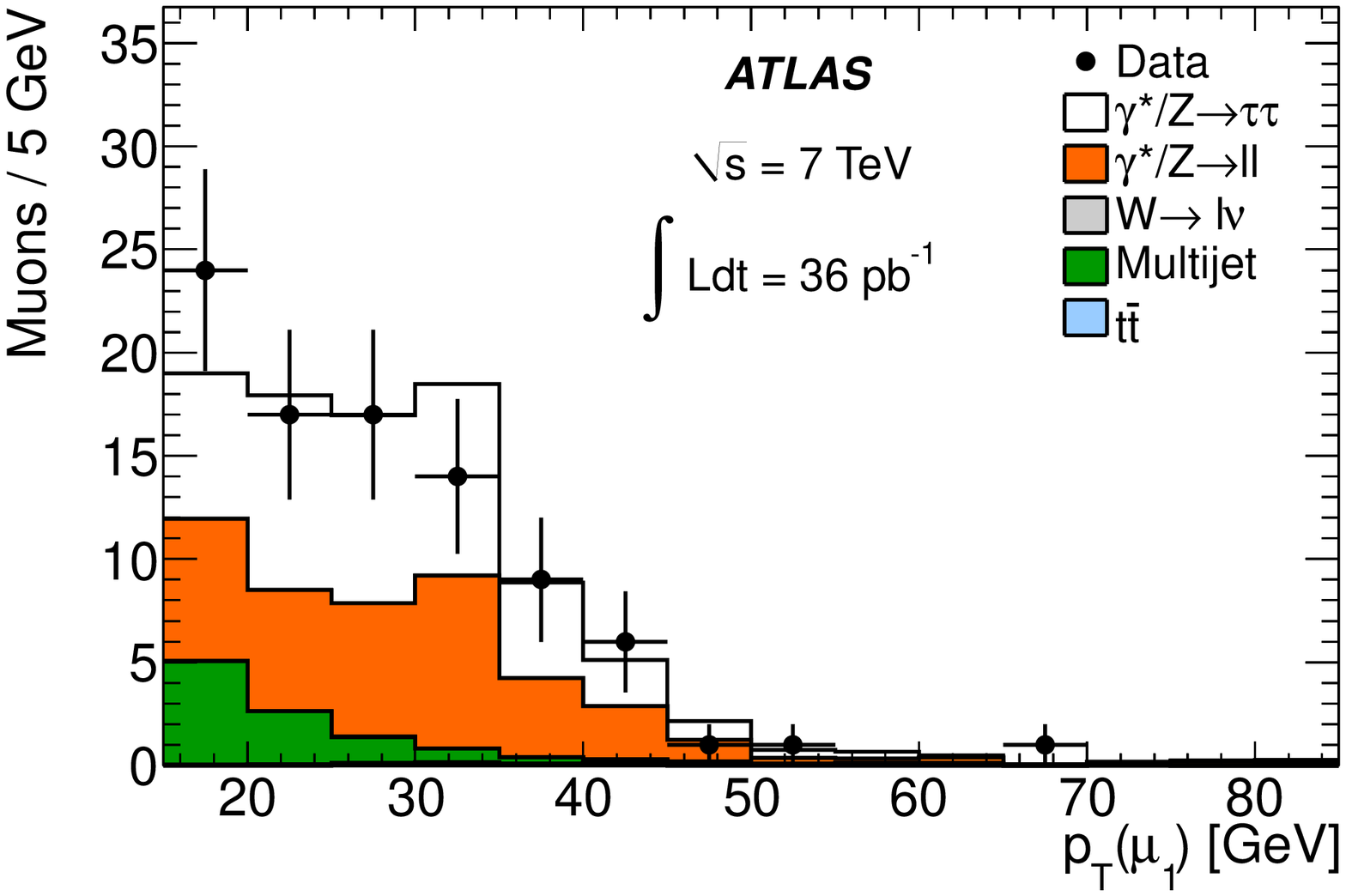}
        \label{fig:mumu_mu_pt}
    }   
    \subfigure[]{
        \includegraphics[width=0.48\textwidth]{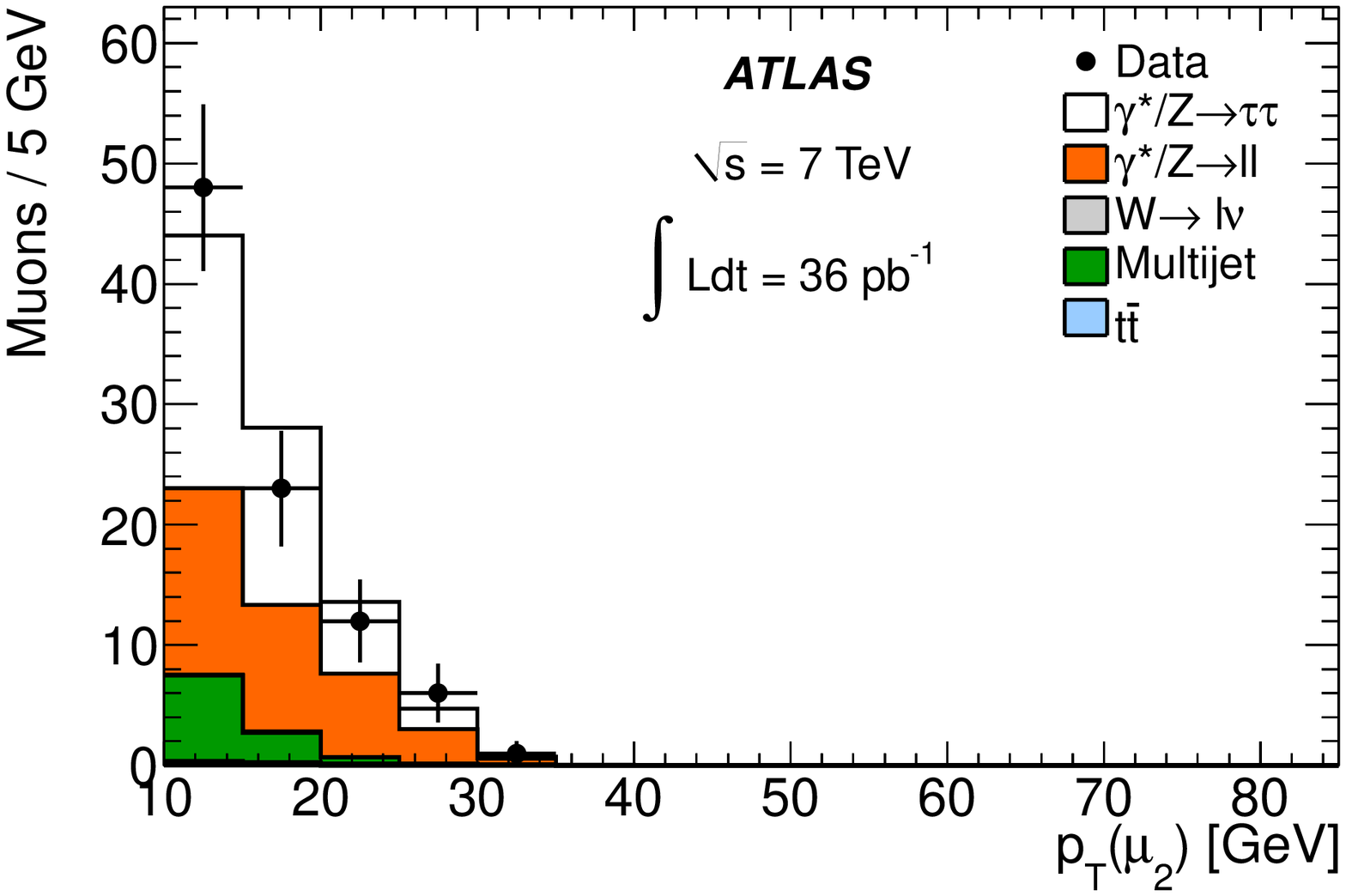}
        \label{fig:mumu_mu2_pt} 
    }  
    \caption{
        Distributions of the $p_\mathrm{T}$ of (a) the leading and
        (b) the subleading muons, for events passing all criteria
        for the $\tau_{\mu}\tau_{\mu}$ final state.
	\label{fig:mumu_kin_var}}
\end{figure}

The following input variables to the BDT training are used:
the difference in azimuthal angle between the two muon candidates ($\Delta \phi(\mu_1,\mu_2)$),
the difference in azimuthal angle between the leading muon candidate and the $\met$ vector~($\Delta \phi(\mu_1 ,\met)$),
the difference in the \pt\ of the two muon candidates ($\pt(\mu_1) - \pt(\mu_2)$), the transverse momentum of the
leading muon candidate ($\pt(\mu_1)$),
and the sum of the absolute transverse impact parameters of the two muon candidates ($| d_0(\mu_1)| + |d_0(\mu_2)|$).
Distributions of these variables for the events that are passed to the BDT
are shown in Figure~\ref{fig:bdt_var}. Differences between data and Monte Carlo
are consistent with the estimated systematic uncertainties, and the agreement is best in the regions most relevant for the signal and background separation.
The sum of the muon transverse impact parameters has the highest discriminating power between the signal and the  $\gamma^*/ Z\rightarrow \mu\mu$ background.
Figure~\ref{fig:BDTscore} shows the distribution of the BDT output. Good agreement between data and MC is observed. 
Events are selected by requiring a BDT output greater than $0.07$.
Cutting on this value gives the best signal significance, and has an efficiency of $0.38 \pm 0.02$.
The visible mass distribution after the full selection except the mass window requirement
can be seen in Figure~\ref{fig:mumu_vis_mass}  and compared to the data.
Figure~\ref{fig:mumu_kin_var} shows the distributions of the $p_\mathrm{T}$ of the two muon candidates passing the full \mumuchan\ selection.

\section{Background estimation}
\label{sec:background_estimation}

In order to determine the purity  of the selected
$Z\rightarrow\tau\tau$ events and the $Z\rightarrow\tau\tau$ production cross section, 
the number of background events passing
the selection criteria  must be estimated.  
The contributions from
the \mbox{$\gamma^{*}/Z\rightarrow\ell\ell$}, \mbox{$t\bar{t}$} and diboson backgrounds
are taken from Monte Carlo simulations, while all other backgrounds are estimated using partially or fully data-driven methods.

\subsection{$W$+jets background}
In the two dileptonic final states, the \mbox{$W\rightarrow\ell\nu$} and \mbox{$W\rightarrow\tau\nu$} backgrounds are found to be 
small, and their contribution is similarly obtained from simulations.
In the two semileptonic final states, where these backgrounds are important, they are 
instead constrained with data by 
obtaining their normalization 
from a \Wboson\ boson-enriched control region. 
This normalization corrects the Monte Carlo for an overestimate of 
the probability for quark and gluon jets produced in association with the \Wboson\
to be misidentified as hadronic $\tau$ decays.
The control region is defined to contain events passing all selection criteria except those
($m_T$, \mbox{$\sum\cos\Delta\phi$}) rejecting the $W$ background. 
This provides a high-purity $W$ sample.
The multijet background contamination in this region is expected to be negligible, while the Monte Carlo estimate of the small \mbox{$\gamma^{*}/Z\rightarrow\ell\ell$} and \ttbar\ 
contribution is subtracted before calculating the normalization factor. The obtained 
normalization factor is 0.73 $\pm$ 0.06 (stat) for the $\tau_{\mu}\tau_{h}$ final state and 0.63 $\pm$ 0.07 (stat) for the $\tau_{e}\tau_{h}$ final state.

\subsection{$\gamma^* / Z \rightarrow \mu\mu$ background}

The most important electroweak background to the $\tau_\mu \tau_\mu$ final state comes from $\gamma^* / Z \rightarrow \mu\mu$ events. 
The normalization of the Monte Carlo simulation is cross-checked after the dimuon selection, for events with invariant masses between 25 GeV and 65 GeV. 
In this region, the $\gamma^* / Z \rightarrow \mu\mu$ process is dominant and is expected to contribute to 
over 94\% of the selected events. The expected backgrounds arising from other electroweak processes are subtracted and
the multijet contribution estimated using a data-driven method described later in this section.
The number of $\gamma^* / Z \rightarrow \mu\mu$ events in the selected mass window is consistent between Monte
Carlo and data within the uncertainties of  $\sim8\%$ (to be compared with a 7\% difference in rate). 
Therefore no correction factor is applied to the $\gamma^* / Z \rightarrow \mu\mu$ 
Monte Carlo prediction.

\begin{table*}[htb]
  \caption{
    Expected number of events per process and number of events  observed in data 
    for an integrated luminosity of $36$~pb$^{-1}$, after the full selection. The background estimates have been obtained as described in Section~\ref{sec:background_estimation}. The quoted uncertainties are statistical only. 
  }
  \centering
    \begin{ruledtabular}
      \begin{tabular}{l c c c c}
        & $\tau_{\mu}\tau_{h}$ &  $\tau_{e}\tau_{h}$  &   $\tau_{e}\tau_{\mu}$  &  $\tau_{\mu}\tau_{\mu}$   \\
	\hline
	$\gamma^*/\Zll$               & $11.1 \pm 0.5$ & $6.9 \pm 0.4$  & $ 1.9 \pm 0.1 $ & $ 36 \pm 1 $ \\           
	$\Wln$               & $9.3 \pm 0.7$  &  $4.8 \pm 0.4$ & $ 0.7 \pm 0.2 $ & $ 0.2  \pm 0.1 $ \\
	$\Wtau$              & $3.6 \pm 0.8$  & $1.5 \pm 0.4$ & $ < 0.2 $ & $ < 0.2 $ \\
	$t\bar{t}$           & $1.3 \pm 0.1$  & $1.02 \pm 0.08$  & $ 0.15 \pm 0.03 $ & $ 0.8 \pm 0.1 $ \\
	Diboson              & $0.28 \pm 0.02$& $0.18 \pm 0.01$ & $ 0.48 \pm 0.03 $ & $ 0.13 \pm 0.01 $ \\
	Multijet             & $24 \pm 6$     & $23 \pm 6$  & $ 6 \pm 4 $ & $ 10 \pm 2 $ \\
	\hline
	$\gamma^*/\Ztau$              & $186 \pm 2$    & $98 \pm 1$ & $ 73\pm 1 $ & $ 44 \pm 1 $ \\
	\hline
	Total expected events    & $235 \pm 6$     & $135 \pm 6$  & $ 82 \pm 4 $ & $ 91 \pm 3 $ \\
	\hline
	$N_{\mathrm{obs}}$ &$ 213 $       & $ 151 $&$ 85 $&$ 90 $ \\
      \end{tabular}
    \end{ruledtabular}
  \label{tab:events}
\end{table*}

\subsection{Multijets}
The multijet background estimation is made by employing data-driven methods in all final states. 
In the $\tau_{e}\tau_{\mu}$, $\tau_{\mu}\tau_{h}$ and $\tau_{e}\tau_{h}$ final states, a multijet enriched control region 
is constructed by requiring the two candidate $\tau$ decay products to have the same sign. 
The ratios of events where the  decay products have the opposite sign to those where they have the same sign $R_{OS/SS}$  is then measured in a separate pair of control regions where the lepton isolation requirement is inverted. 
Electroweak backgrounds in all three control regions are subtracted using Monte Carlo simulations. For the same-sign control regions of the semileptonic final states, the $W$ normalization factor is recomputed using a new $W$ control region identical to that described above, except for having the same-sign requirement applied. The reason is that the sign requirement changes the relative fraction of quark- and gluon-induced jets leading to different $\tau$ 
 misidentification probabilities. 
The following values of $R_{OS/SS}$ are obtained:

\begin{tabular}{cl}
     $1.07 \pm 0.04~(\mathrm{stat}) \pm 0.04~(\mathrm{syst})$ & $\tau_{\mu}\tau_{h}~\mathrm{final~state}$ \\
     $1.07 \pm 0.07~(\mathrm{stat}) \pm 0.07~(\mathrm{syst})$ & $\tau_{e}\tau_{h}~\mathrm{final~state}$\\
     $1.55 \pm 0.04~(\mathrm{stat}) \pm 0.20~(\mathrm{syst})$ & $\tau_{e}\tau_{\mu}~\mathrm{final~state}$.\\
\end{tabular}

The $R_{OS/SS}$ ratios measured in non-isolated events are applied to the same-sign isolated events in order to estimate the multijet contribution to the signal region.
The multijet background is estimated after the full selection 
in the two semileptonic final states, and after the dilepton
selection in the  $\tau_{e}\tau_{\mu}$ final state, due to limited statistics.
The efficiency of the remaining selection criteria is obtained from the same-sign non-isolated 
control region.

This method assumes that the $R_{OS/SS}$ ratio is the same for non-isolated and isolated
leptons. The measured variation of this ratio as a function of the isolation requirements
is taken as a systematic uncertainty.

The multijet background to the $\tau_{\mu}\tau_{\mu}$ final state is estimated   
in a control region defined as applying the full selection, but requiring the subleading muon candidate to fail the isolation selection criteria.
A scaling factor is then calculated in a separate pair of control regions, obtained by requiring that  
the leading muon candidate fails the isolation selection and that the subleading muon candidate either fails or passes it.
This scaling factor is further corrected for the correlation between the isolation variables for the two muon candidates. 
The multijet background in the signal region is finally obtained from the number of events in 
the primary control region scaled by the corrected scaling factor.

\subsection{Summary}
Table~\ref{tab:events} shows the estimated number of background events per process for all channels. The full selection described in Section~\ref{sec:event_selection} has been applied. Also shown are the expected number of signal events, as well as the total number of events observed in data in each channel after the full selection.

\section{cross section calculation}
\label{sec:azcz}

The measurement of the cross sections is obtained using the formula

\begin{equation}\label{Eq:tot_incl_xsec}
  \sigma(Z \ra \tau \tau) \times \mathrm{B}  = \frac{N_{\mathrm{obs}} - N_{\mathrm{bkg}}}{A_{Z} \cdot  C_{Z} \cdot  \mathcal{L}}
\end{equation}

\noindent
where $N_{\mathrm{obs}}$ is the number of observed events in data, $N_{\mathrm{bkg}}$ is the number of estimated background events, 
B is the branching fraction for the channel considered and $\mathcal{L}$ denotes the integrated luminosity for the final state of interest.
$C_{Z}$ is the correction factor that accounts for the efficiency of triggering, reconstructing and identifying the \Ztau\ events within 
the fiducial regions, defined as:
\\

\noindent\begin{tabular}{ll}
$\tau_{\mu}\tau_{h}$ final state: & \\ 
Muon &$ p_{\mathrm{T}} > 15 \mbox{ GeV, } |\eta| < 2.4 $\\
Tau &$  p_{\mathrm{T}} > 20 \mbox{ \GeV, } |\eta| < 2.47,$Ê\\
&$ \mbox{excluding } 1.37 < |\eta| < 1.52 $\\
Event &$ \Sigma \cos \Delta \phi > -0.15 , \mT < 50\mbox{ \GeV, } $ \\
& $\mbox{\mvis\ within [35, 75] \GeV} $\\
\end{tabular}
\begin{tabular}{ll}
 $\tau_{e}\tau_{h}$  final state: &\\
Electron &$ \mbox{\ET} > 16 \mbox{ \GeV, } \, |\eta| < 2.47,$Ê\\
&$ \mbox{excluding } 1.37 < |\eta| < 1.52 $\\
Tau &$  p_{\mathrm{T}} > 20 \mbox{ \GeV, } |\eta| < 2.47,$Ê\\
&$ \mbox{excluding } 1.37 < |\eta| < 1.52 $\\
Event &$ \Sigma \cos \Delta \phi > -0.15 , \mT < 50\mbox{ \GeV, } $\\
&$\mbox{\mvis\ within [35, 75] \GeV} $\\
\end{tabular}
\begin{tabular}{ll}
 $\tau_{e}\tau_{\mu}$ final state: &\\
Electron &$ \mbox{\ET} > 16 \mbox{ \GeV, } \, |\eta| < 2.47,$\\
&$ \mbox{excluding } 1.37 < |\eta| < 1.52 $\\
Muon & $ p_{\mathrm{T}} > 10 \mbox{ GeV, } |\eta| < 2.4 $\\
Event & $ \Sigma \cos \Delta \phi > -0.15$,\\
& $ \mbox{\mvis\ within [25, 80] \GeV} $\\
\end{tabular}
\begin{tabular}{ll}
$\tau_{\mu}\tau_{\mu}$ final state: &\\
Leading muon & $ p_{\mathrm{T}} > 15 \mbox{ GeV, } |\eta| < 2.4 $\\
Subleading muon & $ p_{\mathrm{T}} > 10 \mbox{ GeV, } |\eta| < 2.4 $\\
Event & $ \mbox{\mvis\ within [25, 65] \GeV}$\\
\end{tabular} 
\\

\noindent
The $C_{Z}$ factor is determined as the ratio between the number of events passing the entire analysis 
selection after full detector simulation and the 
number of events in the fiducial region at generator level.
The four-momenta of electrons and muons are calculated including photons radiated within 
a cone of size $\Delta R = 0.1$. The four-momenta of the $\tau$ candidates are defined by including photons radiated by both 
the $\tau$ leptons and their decay products within a cone of size $\Delta R = 0.4$.
By construction $C_{Z}$ accounts for  migrations from outside of the acceptance. 
The correction by the $C_{Z}$ factor provides the cross section within the fiducial region of each measurement

\begin{equation}\label{Eq:fid_xsec}
  \sigma^{\mathrm{fid}}(Z \ra \tau \tau) \times \mathrm{B} = \frac{N_{\mathrm{obs}} - N_{\mathrm{bkg}}}{C_{Z} \cdot \mathcal{L}},
\end{equation}

\noindent
which is independent of the extrapolation
procedure to the full phase space, and therefore is less affected by theoretical uncertainties in the modeling of the \Zboson~production. 

The acceptance factor $A_{Z}$ allows the extrapolation of $\sigma^{\mathrm{fid}}$ to the total cross section, defined by Eq.~\ref{Eq:tot_incl_xsec}.
The $A_{Z}$ factor is determined from Monte Carlo as
the ratio of events at generator level whose $\tau \tau$ invariant mass, before final state radiation (FSR), lies within the mass window [66, 
116]~\GeV, and the number of events at generator level that fall within the fiducial regions defined above. 

The $A_{Z}$ factor accounts for events that migrate from outside the invariant mass window into the fiducial selection criteria. 
The central values for $A_{Z}$ and $C_{Z}$ are determined using a \textsc{pythia} Monte Carlo sample generated with the modified LO parton distribution 
functions 
(PDFs) MRSTLO*~\cite{mrst} and the corresponding  ATLAS MC10 tune~\cite{MC10}. 

\section{Systematic uncertainties}
\label{sec:systematics}

\subsection{Systematic uncertainty on signal and background predictions}
\label{sec:systematic_pred} 

\paragraph{Efficiency of lepton trigger, identification, and isolation}
As described in Sections \ref{sec:data_samples} and \ref{sec:event_selection},
the efficiency of the lepton trigger, reconstruction, identification and isolation requirements are each 
measured separately in data, and the corresponding Monte Carlo efficiency for each step is 
corrected to agree with the measured values.
These corrections are applied to all relevant Monte Carlo samples used for this study. 
Uncertainties on the corrections arise both from statistical and systematic uncertainties 
on the efficiency measurements. 

For the electrons, when estimating the effect of these uncertainties on the signal yield and on 
the background predictions for each final state,  the uncertainties of the individual measurements 
are conservatively treated as uncorrelated to each other and added in quadrature. 
The largest contribution to the electron efficiency uncertainty comes from the identification 
efficiency for low-$\et$ electrons, where the statistical uncertainty on the measurement is 
very large. The total electron uncertainty is estimated to be between 5-9\% relative to 
the efficiency, depending on the selection. 

For muons, the uncertainty is determined in the same way as for electrons and is estimated to be 2-4\% relative to the efficiency.

\paragraph{Efficiency of hadronic $\tau$ identification}
The uncertainties on the hadronic $\tau$ reconstruction and identification efficiencies are 
evaluated by varying simulation conditions, such as the underlying event model, the amount of detector material, the hadronic shower model 
and the noise thresholds of the calorimeter cells in the cluster reconstruction. 
These contributions are added in quadrature to obtain the final systematic uncertainty in bins of $\pt{}$ of the $\tau$ 
candidate and independently for the one track and three track $\tau$ candidates and for low ($\leq 2$) and high multiplicity of primary vertices 
in the event. 
The latter categorization is necessary due to the effects of pile-up 
(additional soft interactions in the same bunch crossing
as the interaction that triggered the readout). 
In events with a large number of additional interactions the $\tau$ identification performance worsens, since the discriminating variables are diluted due to the increased activity in the tracker and calorimeters.
The systematic uncertainties are estimated to be around 10\% relative to the efficiency for most cases, varying 
between 9\% and 12\% with the $\tau$ candidate \pt, number of tracks, and number of vertices in the event~\cite{ATLAS-CONF-2011-057}.

\paragraph{Electron and jet misidentification as $\tau$ candidates}
The probability for an electron  or a QCD jet to be misidentified as a hadronic $\tau$ is measured in data. 
The misidentification probability for electrons is determined using an identified $Z \ra ee$ sample where $\tau$ identification is applied to one of the electrons. 
Correction factors are derived for the Monte Carlo misidentification probability for electrons, binned in $\eta$. 
These corrections are applied to $\tau$ candidates matched in simulation 
to a generator-level electron, with the uncertainty on the correction factor taken as the systematic uncertainty. 
The QCD jet misidentification probability is measured in $Z\rightarrow\ell\ell$+jet events. The difference to the Monte Carlo 
prediction for the same selection, added in quadrature with the statistical and systematic uncertainties of the measurement, is taken as the systematic uncertainty.
These corrections are applied to $\tau$ candidates not matched to a generator-level electron. 
The $\tau$ candidate misidentification systematic uncertainties are not applied to the \Wboson\ Monte Carlo samples, as these have been normalized to data to account for the QCD jet misidentification probability. Instead the uncertainty on the normalization is applied, as described later in this section.

\begin{table*}[htb]
  \caption{
    Relative statistical and systematic uncertainties in \% on the total cross section measurement. The electron and muon efficiency terms include the lepton trigger, reconstruction, identification and isolation uncertainties, as described in the text. The last column indicates whether a given systematic uncertainty is treated as  correlated (\checkmark) or uncorrelated (X) among the relevant channels when combining the results, as described in Section~\ref{sec:combination}. For the multijet background estimation method, the uncertainties in the  $\tau_{\mu}\tau_{h}$,  $\tau_{e}\tau_{h}$, and $\tau_{e}\tau_{\mu}$ channels are treated as correlated while the $\tau_{\mu}\tau_{\mu}$ uncertainty is treated as uncorrelated, since a different method is used, as described in Section~\ref{sec:background_estimation}.
  }
  \centering
    \begin{ruledtabular}
      \begin{tabular}{l c c c c c}
        Systematic uncertainty & $\tau_{\mu}\tau_{h}$ &  $\tau_{e}\tau_{h}$  &   $\tau_{e}\tau_{\mu}$  &  $\tau_{\mu}\tau_{\mu}$ & Correlation   \\
	\hline
	Muon efficiency                             &        3.8\% &        --      & 2.2\% & 8.6\% & \checkmark \\
	Muon $d_0$ (shape and scale)                  & --         & --              & --  &  6.2\% & X \\
	Muon resolution \& energy scale         &        0.2\% &      --      & 0.1\% & 1.0\% & \checkmark \\
	Electron efficiency, resolution \&             &        &             & &  &  \\   
	Charge misidentification                          &        -- &        9.6\%      & 5.9\% & -- & \checkmark \\
	$\tau_{h}$ identification efficiency                         &        8.6\% &        8.6\%      & -- & -- & \checkmark \\
	$\tau_{h}$ misidentification                   &        1.1\% &       0.7\%      & -- & -- & \checkmark \\
	Energy scale (e/$\tau$/jets/\met)                &         10\% &         11\%      & 1.7\% & 0.1\%& \checkmark \\
	Multijet estimate method                      &       0.8\% &          2\%      & 1.0\% & 1.7\% & (\checkmark)  \\
	\Wboson\ normalization factor                                  &       0.1\% &       0.2\%      & --  & -- & X \\
	Object quality selection criteria                           &        1.9\% &        1.9\%      & 0.4\% & 0.4\%& \checkmark \\
	pile-up description in simulation                            &       0.4\% &       0.4\%      & 0.5\% & 0.1\%& \checkmark \\
	Theoretical cross section                  &        0.2\% &        0.1\%     &  0.3\%& 4.3\% &  \checkmark  \\
	\hline
	$A_{Z}$ systematics                           &        3\% &        3\%      &  3\%& 4\%& \checkmark  \\
	\hline 
	Total Systematic uncertainty                  &         15\% &         17\%     & 7.3\% & 14\% & \\
	Statistical uncertainty                       &        9.8\% &         12\%     & 13\%  & 23\% & X \\
	Luminosity                                    &        3.4\% &        3.4\%      & 3.4\%  & 3.4\% & \checkmark \\
      \end{tabular}
    \end{ruledtabular}
  \label{tab:syst_xs}
\end{table*}

\paragraph{Energy scale}
The $\tau$ energy scale  uncertainty is estimated by varying the detector geometry, hadronic showering model, underlying 
event model as well as the noise thresholds of the calorimeter cells in the cluster reconstruction in the simulation, and comparing 
to the nominal results~\cite{ATLAS-CONF-2011-057}. 
The electron energy scale is  determined from data by constraining the reconstructed dielectron invariant mass to the well-known $Z\ra ee$ line shape. 
For the central region the linearity and resolution are in addition controlled using $J/\psi \ra ee$ events. 

The jet energy scale uncertainty is evaluated from simulations
by comparing the nominal results to Monte Carlo simulations using alternative detector configurations, alternative hadronic 
shower and physics models, and by comparing the relative response of jets across pseudo-rapidity between data and simulation~\cite{jet_calib}. 
Additionally, the calorimetric component of the \met\ is sensitive to the energy scale, and this uncertainty is evaluated by propagating first the electron energy scale uncertainty into the \met\ calculation and then shifting all topological clusters not associated to electrons according to their uncertainties~\cite{jet_calib}.

The electron, $\tau$ and jet energy scale uncertainties, as well as the calorimetric component of the \met, are all correlated. Their effect is therefore evaluated by simultaneously shifting each up and down by one standard deviation; the jets are not considered in the semileptonic final states, while the $\tau$ candidates are not considered for the dilepton final states.
The muon energy scale, and the correlated effect on the \met, is also evaluated but found to be negligible in comparison with other uncertainties.

\paragraph{Background estimation}
The uncertainty on the multijet background estimation arises from three separate areas.  
Electroweak and \ttbar~backgrounds are subtracted in the control regions and all sources of systematics on these backgrounds are taken 
into account.
Each source of systematic error is varied up and down by one standard deviation and the effect on the final multijet background estimation
is evaluated.

The second set of systematic uncertainties is related to the assumptions of the method used for the $\tau_{e}\tau_{h}$, $\tau_{\mu}\tau_{h}$, and  $\tau_{e}\tau_{\mu}$ final state multijet background estimations, that the ratio of opposite-sign to same-sign events in the signal region is independent of the lepton isolation. 

These systematic uncertainties are evaluated by studying the dependence of $R_{OS/SS}$ on the isolation variables selection criteria and, for the $\tau_e\tau_\mu$ channel, comparing the efficiencies of the subsequent selection criteria in the opposite and same sign regions.
For the estimation of the multijet background in the  $\tau_{\mu}\tau_{\mu}$ final state, the uncertainties due to the correlation between the isolation of the two muon candidates are evaluated by propagating the systematic uncertainties from the subtracted backgrounds into the calculation of the correlation factor.
The third uncertainty on the multijet background estimation arises from the statistical uncertainty on the number of data events in the various control regions. 

The uncertainty on the $W$+jets background estimation method is dominated by the statistical uncertainty on the calculation of 
the normalization  factor in the control region, as described in Section~\ref{sec:background_estimation},
and the energy scale uncertainty.

\paragraph{Muon $d_0$ smearing}
In the  $\tau_{\mu}\tau_{\mu}$ final state, a smearing is applied to the transverse impact parameter of 
the muons with respect to the primary vertex ($d_0$) to match the Monte Carlo resolution with the value observed in data. 
The muon $d_0$ distribution is compared between data and Monte Carlo using a sample of \Zmumu\ events 
and it is found to be well-described by a double Gaussian distribution.
The 20\% difference in width between data and simulation is used to define a smearing function which is applied to the $d_0$ of each simulated muon.
The systematic uncertainty due to the smearing procedure is estimated by varying the widths and the relative weights of the two components of the impact parameter distributions applied to the Monte Carlo, within the estimated uncertainties on their measurement. An additional uncertainty is found for the \Ztau\ signal sample.

\paragraph{Other sources of systematic uncertainty}
The uncertainty on the luminosity is taken to be 3.4\%, as determined in~\cite{ATLAS-CONF-2011-011,lumi:2011}. 
A number of other sources, such as the uncertainty due to the object quality requirements 
on $\tau$ candidates and on jets,
are also evaluated but have a small impact on the total uncertainty.
The Monte Carlo is reweighted so that the distribution of the number of vertices matches that observed in data; the systematic uncertainty from the reweighting procedure amounts to a permille effect.  The lepton resolution and charge misidentification are found to only have a sub-percent effect on $C_{Z}$ and the background predictions. Systematic uncertainties due to a few problematic calorimetric regions, affecting electron reconstruction, are also evaluated and found to a have a very small effect.  The uncertainties on the theoretical cross sections by which the background Monte Carlo samples are scaled are also found to only have a very small impact on the corresponding background prediction, except for the  $\tau_{\mu}\tau_{\mu}$ final state, which has a large electroweak background contamination.

\subsection{Systematic uncertainty on the acceptance}

The theoretical uncertainty on the geometric and kinematic acceptance factor $A_{Z}$ is dominated by the limited knowledge of the proton PDFs and the modeling of the Z-boson production at the LHC. 
The uncertainty due to the choice of PDF set is evaluated by considering the maximal deviation between the acceptance obtained using the default sample and the values obtained by reweighting this sample to the CTEQ6.6 and HERAPDF1.0~\cite{herapdf} PDF sets. 
The uncertainties within the PDF set are determined by using the 44 PDF error eigenvectors available~\cite{cteq66} for the CTEQ6.6 NLO PDF set. The variations are obtained by reweighting the default sample to the relevant CTEQ6.6 error eigenvector. 
The uncertainties due to the modeling of W and Z production are estimated 
using \textsc{mc@nlo} interfaced with \textsc{herwig} for parton showering, with the CTEQ6.6 PDF set and ATLAS MC10 tune and a lower bound on the invariant mass of
60~GeV. Since \textsc{herwig} in association with external generators does not handle $\tau$ polarizations correctly~\cite{herwigtauola}, the acceptance obtained from the 
\textsc{mc@nlo} sample is corrected for this
effect, which is of order 2\% for the $\tau_{e}\tau_{h}$ and $\tau_{\mu}\tau_{h}$ channels, 8\% for the $\tau_{e}\tau_{\mu}$ channel, and 3\% for the $\tau_{\mu}\tau_{\mu}$ channel.
The deviation with respect to the $A_{Z}$ factor obtained using the default sample reweighted to the CTEQ6.6 PDF set central value and with an applied lower bound on the invariant mass of 60~GeV is taken as uncertainty. 
In the default sample the QED radiation is modeled by \textsc{photos}  
which has an accuracy of better than 0.2\%, and therefore has a  negligible uncertainty compared to uncertainties due to PDFs.
Summing in quadrature the various contributions,  total theoretical
uncertainties of 3\% are assigned to $A_{Z}$ for both of the semileptonic and 
the $\tau_{e}\tau_{\mu}$ final states and of 4\% for the $\tau_{\mu}\tau_{\mu}$ final state.

\subsection{Summary of systematics}

The uncertainty on the experimental acceptance $C_{Z}$  is given by the effect of the 
uncertainties described in Section~\ref{sec:systematic_pred} on the signal Monte Carlo, after correction factors have been applied.
For the total background estimation uncertainties, the correlations between the electroweak and \ttbar\ background uncertainties and the multijet background uncertainty, arising from the subtraction of the former in the control regions used for the latter, are taken into account.
The largest uncertainty results from the $\tau$ identification and energy scale uncertainties for the $\tau_{\mu}\tau_{h}$ and $\tau_{e}\tau_{h}$ final states. 
Additionally, in the  $\tau_{e}\tau_{h}$ final state, the uncertainty on  the electron efficiency has a large contribution. 
This is also the dominant uncertainty in the $\tau_{e}\tau_{\mu}$  final state. 
In the $\tau_{\mu}\tau_{\mu}$ final state, the uncertainty due to the muon efficiency is the dominant source, with the muon $d_0$ contribution being important in the background estimate contributions for that channel.
The correlation between the uncertainty on $C_{Z}$  and on  ($N_{\mathrm{obs}}-N_{\mathrm{bkg}}$) is accounted for in obtaining the final uncertainties on the cross section measurements, which are summarized in Table~\ref{tab:syst_xs}.

\section{cross section measurement}
\label{sec:results}

\subsection{Results by final state}

The determination of the cross sections in each final state is performed by using the numbers from the previous sections, provided for reference in Table~\ref{tab:result_overview}, following the method described in Section~\ref{sec:azcz}. 
Table~\ref{tab:input} shows the cross sections measured individually in each of the four final states. Both the fiducial cross sections and the total cross sections for an invariant mass window of  [66,~116]~\GeV\ are shown.

\begin{table}[htdp]
  \caption{The components of the $Z \rightarrow \tau\tau$ cross section calculations for each final state. For $N_{\mathrm{obs}} - N_{\mathrm{bkg}}$ the first uncertainty is statistical and the second systematic. For all other values the total error is given.}
  \begin{center}
    \begin{ruledtabular}
      \begin{tabular}{c c c}
	& $\tau_{\mu}\tau_{h}$ &  $\tau_{e}\tau_{h}$    \\
	\hline
	$N_{\mathrm{obs}}$ &$ 213 $       & $ 151 $   \\
	$N_{\mathrm{obs}}-N_{\mathrm{bkg}}$ &$ 164 \pm 16 \pm 4  $&$  114 \pm 14 \pm 3 $  \\
	$A_{Z}$ &$ 0.117 \pm 0.004$&$ 0.101 \pm 0.003  $   \\
	$C_{Z}$ &$ 0.20  \pm 0.03  $&$  0.12 \pm 0.02  $ \\
	B & $0.2250 \pm 0.0009$& $0.2313 \pm 0.0009$\\
	$\mathcal{L}$ &$ 35.5 \pm 1.2~$pb$^{-1}  $&$  35.7 \pm 1.2~$pb$^{-1}   $  \\
	\hline\hline
	& $\tau_{e}\tau_{\mu}$  &  $\tau_{\mu}\tau_{\mu}$   \\
	\hline
	$N_{\mathrm{obs}}$  &$ 85 $&$ 90 $ \\
	$N_{\mathrm{obs}}-N_{\mathrm{bkg}}$  &$ 76 \pm 10 \pm 1  $&$  43 \pm 10 \pm 3 $ \\
	$A_{Z}$  &$ 0.114 \pm 0.003$&$ 0.156 \pm 0.006  $ \\
	$C_{Z}$  &$ 0.29 \pm 0.02  $&$  0.27 \pm 0.02  $  \\
	B & $0.0620 \pm 0.0002$ & $0.0301 \pm 0.0001$\\
	$\mathcal{L} $ &$ 35.5 \pm 1.2~$pb$^{-1}   $&$  35.5 \pm 1.2~$pb$^{-1}   $  \\
      \end{tabular}
    \end{ruledtabular}
  \end{center}
  \label{tab:result_overview}
\end{table}

\begin{table}[htbp]
  \caption{
The production cross section times branching fraction for the $Z \ra \tau\tau$ process
as measured in each of the four final states, and the combined result. For the fiducial 
cross sections the measurements include also the branching fraction of the $\tau$ to its decay products. The first error is statistical, the second systematic and the third comes from the luminosity.} 
  \centering
  \begin{ruledtabular}
    \begin{tabular}{l c}
      Final State & Fiducial cross section (pb)\\
      \hline
      $\tau_{\mu} \tau_{h}$ & $23 \pm   2  \pm 3  \pm  1 $ \\
      $\tau_{e} \tau_{h}$ & $27 \pm   3  \pm 5  \pm  1 $ \\
      $\tau_{e} \tau_{\mu}$ & $7.5 \pm   1.0  \pm 0.5  \pm  0.3 $ \\
      $\tau_{\mu} \tau_{\mu}$ &$ 4.5 \pm   1.1  \pm 0.6  \pm  0.2$ \\
      \hline
      Final State & Total cross section ($[66, 116]$~\GeV) (nb)\\
      \hline
      $\tau_{\mu} \tau_{h}$ & $0.86 \pm   0.08  \pm 0.12  \pm  0.03$\\
      $\tau_{e} \tau_{h}$ & $1.14 \pm   0.14 \pm 0.20  \pm  0.04$ \\  
      $\tau_{e} \tau_{\mu}$ & $1.06 \pm   0.14 \pm 0.08 \pm  0.04$ \\
      $\tau_{\mu} \tau_{\mu}$ &$ 0.96 \pm   0.22 \pm 0.12  \pm  0.03$ \\ 
      \hline  
      $Z \ra \tau \tau$ & $ 0.97 \pm   0.07  \pm 0.06  \pm  0.03$ \\
    \end{tabular} 
  \end{ruledtabular}
  \label{tab:input} 
\end{table}

\subsection{Combination} 
\label{sec:combination}

The combination of the cross section measurements from the four final states is obtained by using the Best Linear Unbiased Estimate (BLUE) 
method, described in~\cite{Lyons:1988rp,Valassi:2003mu}.   
The BLUE method determines the best estimate of the combined total cross section using a linear combination built from the individual measurements, with an estimate of $\sigma$ that is unbiased and has the smallest possible variance.   
This is achieved by constructing a covariance matrix from the statistical and systematic uncertainties for each individual cross section measurement, while accounting for correlations between the uncertainties from each channel.

The systematic uncertainties on the individual cross sections due to different sources are assumed to either be fully correlated or fully uncorrelated. All systematic uncertainties pertaining to the efficiency and resolution of the various physics objects used in the four analyses - reconstructed electron, muon, and hadronically decaying tau candidates - are assumed to be fully correlated between final states that make use of these objects. No correlation is assumed to exist between the systematic uncertainties relating to different physics objects. Similarly, the systematic uncertainties relating to the triggers used by the analyses are taken as fully correlated for the final states using the same triggers and fully uncorrelated otherwise.  The systematic uncertainty on the energy scale is conservatively taken to be fully correlated between the final states. 

As the multijet background is estimated using the same method in the $\tau_{e}\tau_{\mu}$, $\tau_{\mu}\tau_{h}$, and $\tau_{e}\tau_{h}$ final states, the systematic uncertainty on the method is conservatively treated as fully correlated. 

Finally, the systematic uncertainties on the acceptance are assumed to be completely correlated, as are the uncertainties on the luminosity and those on the theoretical cross sections used for the normalization of the Monte Carlo samples used to estimate the electroweak and $t\bar{t}$ backgrounds. 

This discussion is summarized in Table~\ref{tab:syst_xs} where the last column indicates whether a given source of systematic uncertainty has been treated as correlated or uncorrelated amongst the relevant channels when calculating the combined result.

Individual cross sections and their total uncertainties for the BLUE combination, as well as   
the weights for each of the final states in the combined cross section, together with their pulls, are 
also shown in Table~\ref{tab:xsec_comb_weights}.

Under these assumptions, a total combined cross section of 
\begin{widetext}
  \begin{equation}
    \sigma(Z \ra \tau \tau,~66<m_{\mathrm{inv}}<116~\GeV) = 0.97 \pm 0.07\mbox{ (stat)} \pm 0.06\mbox{ (syst)} \pm 0.03\mbox{ (lumi) nb}
  \end{equation}
\end{widetext}
is obtained from the four final states, $\tau_{\mu}\tau_{h}$, $\tau_{e}\tau_{h}$, $\tau_{e}\tau_{\mu}$, and $\tau_{\mu}\tau_{\mu}$.

\begin{figure}[hbtp]
    \centering
       \includegraphics[width=0.48\textwidth]{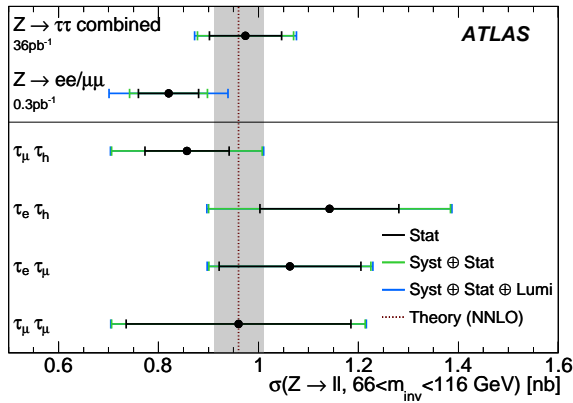}
    \caption{
 The individual cross section measurements by final state, and the combined result. 
 The $Z \ra \ell \ell$ combined cross section measured by ATLAS in the  $Z \ra \mu\mu$ 
 and $Z \ra ee$ final states is also shown for comparison.
 The gray band indicates the uncertainty on the NNLO cross section prediction.       \label{fig:4chanInd}}
\end{figure}

A comparison of the individual cross sections with the combined result is shown in Figure~\ref{fig:4chanInd}, along with  the combined  $Z \ra \ell \ell$ cross section measured in the  $Z \ra \mu\mu$ and $Z \ra ee$ final states by ATLAS~\cite{WZcross}. 
The theoretical expectation of  $0.96 \pm 0.05$ nb for an invariant mass window of
$\mathrm{[66, 116]}$~\GeV\ is also shown. 
The obtained result is compatible with the $Z \ra \tau\tau$ cross section in four final states published recently by the CMS Collaboration~\cite{CMStautau},  $1.00 \pm 0.05\mbox{ (stat)} \pm 0.08\mbox{ (syst)} \pm 0.04\mbox{ (lumi) nb}$, in a mass window of $\mathrm{[60, 120]}$~\GeV. 

\begin{table}[htdp]
  \caption{Individual cross sections and their total uncertainties used in the BLUE
    combination, the weights for each of the final states in the combined   
    cross section, and their pulls.  
    The pull here is defined as the difference between the individual and 
    combined cross sections divided by the uncertainty on this difference.
    The uncertainty on the difference between the measured and combined cross section
    values includes the uncertainties on the cross section
    both before and after the combination, taking all correlations into
    account.}
  \begin{center}
    \begin{ruledtabular}
      \begin{tabular}{c c c c c}   
	&  $\tau_{\mu}\tau_{h}$ &  $\tau_{e}\tau_{h}$ &  $\tau_{e}\tau_{\mu}$ &  $\tau_{\mu}\tau_{\mu}$ \\    
	\hline
	$\sigma_{Z\rightarrow\tau\tau}$ (nb) & 0.86 & 1.14 & 1.06 & 0.96\\  
	Total unc. (nb) & 0.15 & 0.24 & 0.17& 0.25\\  
	Weight & 39.4\% &  7.9\% & 39.0\% & 13.7\% \\  
	Pull   & 1.02  &  -0.76 & -0.68  & 0.06 \\  
      \end{tabular}
      \end{ruledtabular}
  \end{center}
\label{tab:xsec_comb_weights}
\end{table}

\section{Summary}
\label{sec:summary}

A measurement of the $Z \ra \tau \tau$ cross section in proton-proton collisions at $\rts = 7 \TeV$ using the ATLAS detector is presented. Cross sections are measured in four final states, $\tau_{\mu}\tau_{h}$, $\tau_{e}\tau_{h}$, $\tau_{e}\tau_{\mu}$, and $\tau_{\mu}\tau_{\mu}$ 
within the invariant mass range [66, 116]~\GeV. 
The combined measurement is also reported. 
A total combined cross section of $\sigma = 0.97 \pm 0.07\mbox{ (stat)} \pm 0.06\mbox{ (syst)} \pm 0.03\mbox{ (lumi) nb}$
is measured, which is in good agreement with the theoretical expectation and with other measurements.

\begin{acknowledgments}
We thank CERN for the very successful operation of the LHC, as well as the
support staff from our institutions without whom ATLAS could not be
operated efficiently.

We acknowledge the support of ANPCyT, Argentina; YerPhI, Armenia; ARC,
Australia; BMWF, Austria; ANAS, Azerbaijan; SSTC, Belarus; CNPq and FAPESP,
Brazil; NSERC, NRC and CFI, Canada; CERN; CONICYT, Chile; CAS, MOST and
NSFC, China; COLCIENCIAS, Colombia; MSMT CR, MPO CR and VSC CR, Czech
Republic; DNRF, DNSRC and Lundbeck Foundation, Denmark; ARTEMIS, European
Union; IN2P3-CNRS, CEA-DSM/IRFU, France; GNAS, Georgia; BMBF, DFG, HGF, MPG
and AvH Foundation, Germany; GSRT, Greece; ISF, MINERVA, GIF, DIP and
Benoziyo Center, Israel; INFN, Italy; MEXT and JSPS, Japan; CNRST, Morocco;
FOM and NWO, Netherlands; RCN, Norway; MNiSW, Poland; GRICES and FCT,
Portugal; MERYS (MECTS), Romania; MES of Russia and ROSATOM, Russian
Federation; JINR; MSTD, Serbia; MSSR, Slovakia; ARRS and MVZT, Slovenia;
DST/NRF, South Africa; MICINN, Spain; SRC and Wallenberg Foundation,
Sweden; SER, SNSF and Cantons of Bern and Geneva, Switzerland; NSC, Taiwan;
TAEK, Turkey; STFC, the Royal Society and Leverhulme Trust, United Kingdom;
DOE and NSF, United States of America.

The crucial computing support from all WLCG partners is acknowledged
gratefully, in particular from CERN and the ATLAS Tier-1 facilities at
TRIUMF (Canada), NDGF (Denmark, Norway, Sweden), CC-IN2P3 (France),
KIT/GridKA (Germany), INFN-CNAF (Italy), NL-T1 (Netherlands), PIC (Spain),
ASGC (Taiwan), RAL (UK) and BNL (USA) and in the Tier-2 facilities
worldwide.

\end{acknowledgments}

\bibliography{biblio}
\onecolumngrid

\clearpage

\begin{flushleft}
{\Large The ATLAS Collaboration}

\bigskip

G.~Aad$^{\rm 48}$,
B.~Abbott$^{\rm 111}$,
J.~Abdallah$^{\rm 11}$,
A.A.~Abdelalim$^{\rm 49}$,
A.~Abdesselam$^{\rm 118}$,
O.~Abdinov$^{\rm 10}$,
B.~Abi$^{\rm 112}$,
M.~Abolins$^{\rm 88}$,
H.~Abramowicz$^{\rm 153}$,
H.~Abreu$^{\rm 115}$
% ...
\bigskip

% ...
$^{10}$ Institute of Physics, Azerbaijan Academy of Sciences, Baku, Azerbaijan\\
$^{11}$ Institut de F\'isica d'Altes Energies and Universitat Aut\`onoma  de Barcelona and ICREA, Barcelona, Spain\\
% ...
$^{48}$ Fakult\"{a}t f\"{u}r Mathematik und Physik, Albert-Ludwigs-Universit\"{a}t, Freiburg i.Br., Germany\\
$^{49}$ Section de Physique, Universit\'e de Gen\`eve, Geneva, Switzerland\\
% ...
$^{88}$ Department of Physics and Astronomy, Michigan State University, East Lansing MI, United States of America\\
% ...
$^{111}$ Homer L. Dodge Department of Physics and Astronomy, University of Oklahoma, Norman OK, United States of America\\
$^{112}$ Department of Physics, Oklahoma State University, Stillwater OK, United States of America\\
% ...
$^{115}$ LAL, Univ. Paris-Sud and CNRS/IN2P3, Orsay, France\\
% ...
$^{118}$ Department of Physics, Oxford University, Oxford, United Kingdom\\
% ...
$^{153}$ Raymond and Beverly Sackler School of Physics and Astronomy, Tel Aviv University, Tel Aviv, Israel\\
% ...
\end{flushleft}



\end{document}